\documentclass{article}

\usepackage{arxiv}

\usepackage[utf8]{inputenc} 
\usepackage[T1]{fontenc}    
\usepackage{hyperref}       
\usepackage{url}            
\usepackage{booktabs}       
\usepackage{amsfonts}       
\usepackage{nicefrac}       
\usepackage[nopatch=footnote]{microtype}      
\usepackage{lipsum}		
\usepackage{graphicx}
\usepackage{natbib}
\usepackage{doi}
\usepackage{multicol}
\usepackage{graphicx} 
\usepackage{hyperref} 
\usepackage{booktabs}  
\usepackage{blindtext}
\usepackage{amsmath}
\usepackage{amsfonts}
\usepackage{siunitx}
\usepackage{colortbl}
\usepackage{xcolor}
\usepackage{multirow}
\usepackage{multicol}
\usepackage{todonotes}
\usepackage{caption}
\usepackage{subcaption}
\usepackage{float}
\usepackage{longtable}
\usepackage{threeparttable}
\usepackage{pdflscape}
\usepackage[parfill]{parskip}
\usepackage{amssymb}
\usepackage{pifont}
\usepackage{setspace}

\newcommand{\cmark}{\text{\ding{51}}}
\newcommand{\xmark}{\text{\ding{55}}}

\title{Machine learning surrogates for efficient hydrologic modeling: Insights from stochastic simulations of managed aquifer recharge}

\author{Timothy Dai \\
	Department of Computer Science\\
	Stanford University\\
	Stanford, CA, USA \\
	\And
	\href{https://orcid.org/0000-0002-5982-6064}{\includegraphics[scale=0.06]{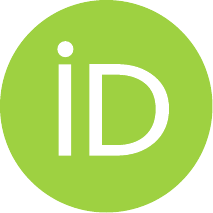}\hspace{1mm}Kate Maher} \\
	Department of Earth System Science\\
	Stanford University\\
	Stanford, CA, USA \\
	\AND
	\href{https://orcid.org/0000-0002-5982-6064}{\includegraphics[scale=0.06]{orcid.pdf}\hspace{1mm}Zach Perzan\thanks{Corresponding author: \texttt{zach.perzan@unlv.edu}.}} \\
	Department of Geoscience \\
	University of Nevada, Las Vegas \\
	Las Vegas, NV, USA \\
}

\date{}

\renewcommand{\headeright}{}
\renewcommand{\undertitle}{}
\renewcommand{\shorttitle}{Manuscript accepted in \emph{Journal of Hydrology}}

\hypersetup{
pdftitle={Machine learning surrogates for efficient hydrologic modeling: Insights from stochastic simulations of managed aquifer recharge},
pdfauthor={Timothy Dai, Kate Maher, Zach Perzan},
pdfkeywords={First keyword, Second keyword, More},
}

\begin{document}
\maketitle

\noindent\fbox{
\begin{minipage}{\textwidth}
\paragraph{Note} This is the accepted version of a manuscript published in the \emph{Journal of Hydrology}. The version of record is:
\vskip 10pt
Dai, T., Maher, K., and Perzan, Z. (2025). Machine learning surrogates for efficient hydrologic modeling: Insights from stochastic simulations of managed aquifer recharge. \emph{Journal of Hydrology}, 652, 132606. doi:\href{https://doi.org/10.1016/j.jhydrol.2024.132606}{10.1016/j.jhydrol.2024.132606}
\end{minipage}}

\vskip 20pt

\begin{abstract}
	Process-based hydrologic models are invaluable tools for understanding the terrestrial water cycle and addressing modern water resources problems. However, many hydrologic models are computationally expensive and, depending on the resolution and scale, simulations can take on the order of hours to days to complete. While techniques such as uncertainty quantification and optimization have become valuable tools for supporting management decisions, these analyses typically require hundreds of model simulations, which are too computationally expensive to perform with a process-based hydrologic model. To address this gap, we assess a hybrid modeling workflow in which a process-based model is used to generate an initial set of simulations and a machine learning (ML) surrogate model is then trained to perform the remaining simulations required for downstream analysis. As a case study, we apply this workflow to simulations of variably saturated groundwater flow at a prospective managed aquifer recharge (MAR) site. We compare the accuracy and computational efficiency of several ML architectures, including deep convolutional networks, recurrent neural networks, vision transformers, and networks with Fourier transforms. Our results demonstrate that ML surrogate models can achieve under 10\% mean absolute percentage error and yield order-of-magnitude runtime savings over process-based models. Building on these findings, we examine the impacts of key modeling choices on surrogate model accuracy and efficiency. Results show that a normalized loss function improves training stability, while min-max data normalization can significantly reduce error up to a factor of 10 when compared to other treatments such as Z-score and no normalization. Downsampling input features using an autoencoder also decreases memory requirements by training with tensors 4\% their original size. By reducing computational costs and memory demands, ML surrogates offer a practical pathway to scale up complex hydrologic analyses in support of sustainable groundwater management.
\end{abstract}


\section{Introduction}
Future water security hinges on effective water resources management. Process-based hydrologic and reactive transport models, which simulate water flow and solute transport \citep{xu_review_2004, devia_review_2015, steefel_reactive_2015}, are frequently used to forecast future changes in the hydrologic cycle and to optimize water management strategies. When paired with uncertainty quantification and parameter estimation methods, these models can support decision-making in diverse scenarios such as flood forecasting, contaminant transport, nutrient availability, and managed aquifer recharge (MAR). These techniques often require hundreds of model simulations in which uncertain input parameters are varied randomly from one simulation to the next. While each individual simulation is deterministic, producing one fixed output for a given set of input parameters, the ensemble of simulations is stochastic because of the random variations in model input.

However, process-based simulations can be computationally expensive to perform. In particular, integrated hydrologic models --- codes that simulate coupled surface and subsurface water flow --- pose several numerical challenges \citep{maxwell_review_2014}, such as solving highly nonlinear governing equations and ensuring flux and pressure continuity at the interface between the surface and subsurface \citep{kollet_integrated_2006}. In addition, to perform simulations at the scale and resolution required for locally relevant policy decisions, model domains must resolve fine-scale heterogeneity in hydraulic conductivity, topography, soil texture, and other parameters \citep{wood_hyperresolution_2011}. These sharp contrasts often result in steep pressure gradients that can cause iterative numerical solvers to slow down or fail to converge. As a result, performing stochastic simulations with integrated hydrologic codes requires large computational resources, even on state-of-the-art supercomputers.

To bridge this gap, recent studies have proposed a hybrid modeling workflow that aims to overcome the computational demands of traditional process-based hydrologic models by capitalizing on the computational efficiency and versatility of modern machine learning (ML) models \citep{asher_review_2015, tran_development_2021, maxwell_physics-informed_2021}. In this workflow, an initial batch of simulations is generated using a process-based hydrologic model. An ML surrogate is then trained on this initial set of simulations and used to perform the remaining model runs. For example, \citet{maxwell_physics-informed_2021} evaluated the accuracy of two- and three-dimensional convolutional surrogate models for simulating overland flow in a simple tilted V catchment. This benchmark test problem \citep{digiammarco_conservative_1996} simulates 2D surface water flow through a simplified domain that consists of two inclined planes that converge in a central channel. Across simulations, the authors varied six uncertain input parameters related to rainfall and surface runoff, all of which were assumed to be spatially homogeneous across the domain. Of the six architectures tested, those that were provided with explicit temporal information during training exhibited the best performance, though accuracy decreased when the simulations in the test set contained different parameter distributions than those in the training set \citep{maxwell_physics-informed_2021}. Building off of this work, \citet{tran_development_2021} trained recurrent and highway networks \citep{wang_predrnn_2018} to emulate surface-subsurface flows within two watersheds spanning $\SIrange{600}{1236}{km^2}$. As compared to a process-based hydrologic model, the emulator exhibits low ($<$$0.1$) relative bias for streamflow and total groundwater storage and can produce simulations 42 times faster.

While these early studies highlight the potential for ML surrogate models to accelerate hydrologic simulations, key knowledge gaps remain in their application to water resources problems. First, the number of process-based simulations required to train a high-fidelity surrogate model remains unclear. Second, when generating a small number of simulations, the added cost of training an ML model may outweigh any speed gained by the trained surrogate, making it more efficient to perform simulations exclusively with a process-based model. Thus, the threshold number of simulations beyond which a hybrid workflow becomes more efficient than process-based modeling, in terms of both computational time and memory, remains poorly understood. Third, the impacts of key modeling choices, such as loss functions and data normalization techniques, on the accuracy and robustness of ML surrogates have yet to be systematically evaluated. Importantly, each of these knowledge gaps may vary across different surrogate model architectures.

In this paper, we investigate these three knowledge gaps and present practical considerations for using ML surrogates to perform stochastic hydrologic simulations. As a case study, we focus on 3D simulations of variably saturated groundwater flow during managed aquifer recharge (MAR), the intentional replenishment of depleted aquifers through recharge basins and injection wells. We train seven different surrogate model architectures to reproduce output from a parallel, integrated hydrologic code (ParFlow-CLM). We then investigate the impacts of training set size, data preprocessing techniques, and choice of loss function on surrogate model performance. We also evaluate the runtime efficiency of each surrogate architecture as compared to the process-based hydrologic code and examine techniques to increase memory efficiency, such as temporal and spatial downsampling of high-dimensional inputs using an autoencoder.

\section{Material and Methods}\label{sec:methods}

In this section, we formalize the hybrid modeling workflow (\S \ref{subsec:propose}) and introduce the case study (\S \ref{subsec:casestudy}). Next, we describe the process-based hydrologic code (\S \ref{subsec:parflow}) and the surrogate model architectures (\S \ref{subsec:ml_models}). We then describe data preprocessing (\S \ref{subsec:data_preproc}) and detail the experimental setup for training and evaluating each ML surrogate (\S \ref{subsec:exp_setup}). We conclude with a summary of experiments, enumerating the ML models trained to evaluate the effects of data normalization, loss function, and training set size on hydrologic surrogate model accuracy and the trade-offs in runtime (\S \ref{subsec:experiments}).

\subsection{The hybrid workflow}\label{subsec:propose}

\begin{figure}
    \centering
    \includegraphics[width=6.5in]{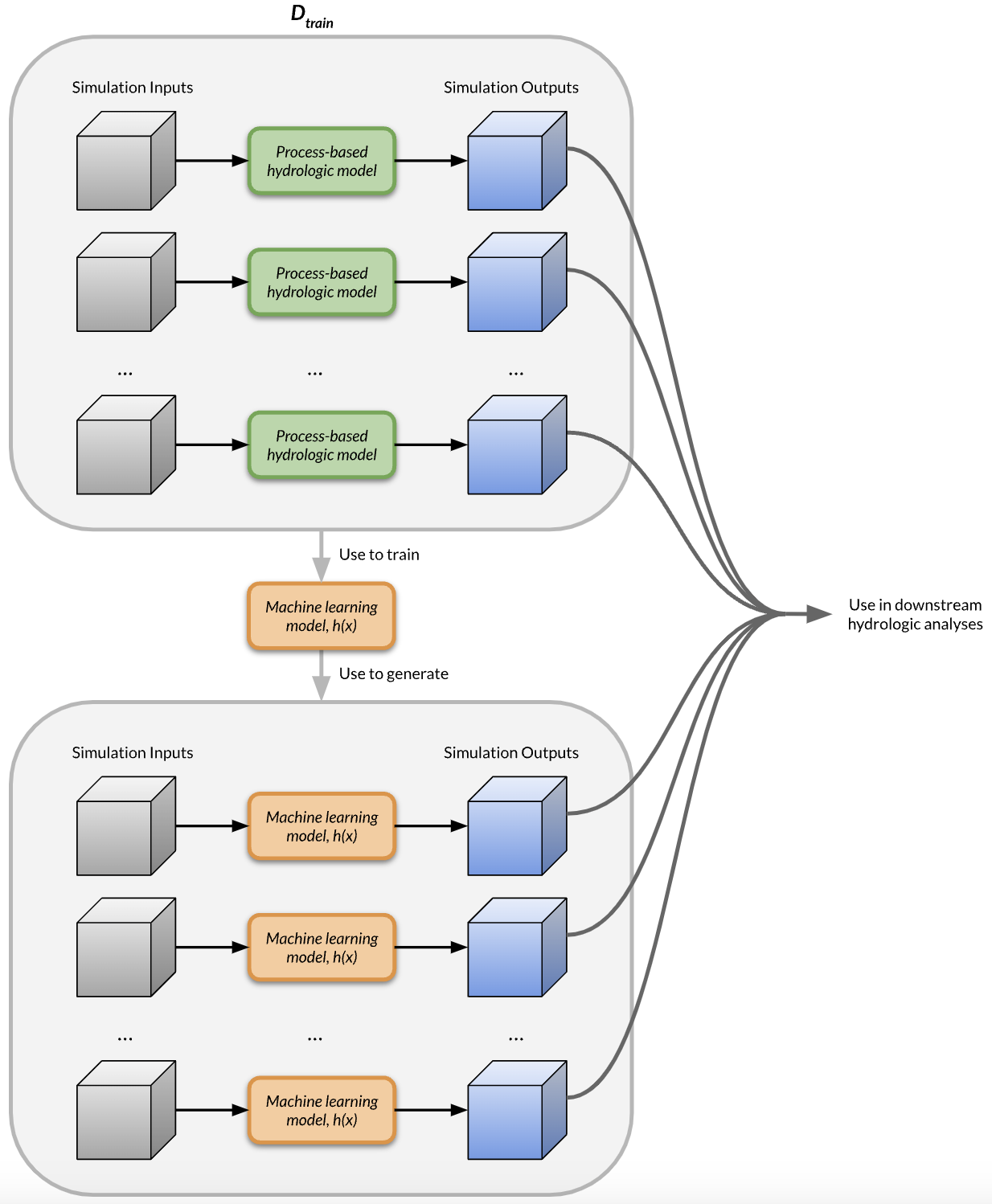}
    \caption{Diagram of the proposed hybrid framework. The process-based hydrologic model generates a training set ($\mathcal D_\text{train}$), which is then used to train and test an ML model. Subsequently, the trained ML model is used to generate the remaining hydrologic simulations.}
    \label{fig:framework_diag}
\end{figure}

The goal of this hybrid workflow is to harness the fast output rate of ML models to efficiently generate a large volume of simulations for downstream hydrologic analysis. The task is to execute $N$ simulations, generating $N$ simulation outputs from $N$ distinct sets of simulation inputs. As illustrated in Figure \ref{fig:framework_diag}, we employ a traditional process-based hydrologic model (ParFlow-CLM) to generate a training set $\mathcal D_\text{train}$ comprising simulation input--output pairs. Subsequently, an ML surrogate model is trained on these examples and used to generate simulation outputs for the remaining $N - |\mathcal D_\text{train}|$ simulations, where $|\mathcal D_\text{train}|$ is the number of examples in $\mathcal D_\text{train}$. In sum, this framework produces $N$ simulations by generating $|\mathcal D_\text{train}|$ simulations with a traditional process-based hydrologic model and $N-|\mathcal D_\text{train}|$ simulations with an ML model. In a hydrologic context, these simulation outputs could aid in a multitude of downstream analyses, including uncertainty quantification, parameter estimation, and resource optimization.

\subsection{The case study: Managed aquifer recharge site}\label{subsec:casestudy}

We apply this hybrid workflow to a single case study in which hydrologic simulations are used to evaluate recharge efficiency at a prospective managed aquifer recharge (MAR) site. MAR is a water management strategy whereby land managers deliberately flood parcels of land or inject water into wells to replenish underlying aquifers. Recent work has shown that MAR is a valuable tool for mitigating groundwater depletion in many regions of the globe \citep{dillon_sixty_2019,sprenger_inventory_2017}, but evaluating prospective recharge sites remains challenging as subsurface heterogeneity can significantly impact recharge rates \citep{knight_airborne_2022,perzan_controls_2023}. Hydrologic and reactive transport simulations are effective tools for quantifying outcomes of interest at prospective MAR sites, such as changes in groundwater storage \citep[e.g., ][]{ganot_managed_2018} or the fate of subsurface contaminants \citep[e.g., ][]{perzan_transport_2024}. However, generating these simulations comes with a notable computational expense, as the time required for a single simulation can range from hours to days depending on model scale and resolution. Thus, MAR site selection is a logical case study that could benefit from a computationally efficient hybrid modeling workflow.

In this case study, we focus on a prospective MAR site in Tulare County, California, that measures $\SI{800}{m}\times\SI{400}{m}$. The site is underlain by a \SI{45}{m} thick vadose zone that consists of interbedded layers of sand, silt and clay. The unconfined aquifer extends to $\SI{150}{m}$ depth, below which the Corcoran Clay, a regional aquitard, limits vertical flow \citep{mid-kaweah_gsa_groundwater_2019}. While this study site has never been used for MAR, the land manager would like to evaluate the site's suitability for future recharge operations. Thus, the goal of these simulations is to quantify the increase in saturated zone storage that would occur within two years of inundating the site with a hypothetical $\SI{0.8}{m}$ recharge event.

To evaluate recharge efficiency at this site, \citet{perzan_controls_2023} performed several hundred stochastic simulations of MAR using ParFlow-CLM (\S\ref{subsec:parflow}). The model domain is discretized into a $120 \times 80 \times 25$ rectilinear grid ($\SI{1200}{m}\times \SI{800}{m} \times \SI{150}{m}$ in the \textit{x}, \textit{y}, and \textit{z} directions, respectively), utilizing a uniform horizontal cell size of $\SI{10}{m}$ and variable cell thicknesses in the $z$ direction ($\SI{1.02}{m}$ at the surface and $\SI{9}{m}$ at $\SI{90}{m}$ depth). To limit boundary effects, the model domain extends \SI{200}{m} beyond the edges of the recharge site in the $x$ and $y$ directions. Specified head boundaries corresponding to the depth to the water table ($\SI{45}{m}$) are applied on the sides of the domain. A no-flow boundary is implemented on the bottom of the domain and variable forcing is applied to the top of the domain based on the modeling stage, as explained at the end of this section.

The domain is parameterized through a stochastic geophysical-geostatistical workflow \citep{perzan_controls_2023}. Because subsurface heterogeneity can exert a dominant control on recharge efficiency, we parameterize the domain using a subsurface electrical resistivity model derived from a towed transient electromagnetic (tTEM) survey of the site. The tTEM system is a hydrogeophysical tool that can map meter-scale variations in electrical resistivity, a property directly related to sediment lithology. We first generate 600 gridded realizations of the tTEM-derived resistivity model using sequential Gaussian simulation, a stochastic technique for populating a rectilinear grid with a random variable. We then compare vertical profiles of electrical resistivity to collocated profiles of sediment type collected through cone penetration tests. Using the bootstrapping procedure of \citet{knight_mapping_2018}, we transform each resistivity realization to a realization of coarse fraction, a metric that describes the fraction of coarse-grained (sand and gravel) sediment within each grid cell. The remainder of sediment within the cell is assumed to be fine-grained (silt and clay). For example, a cell with a coarse fraction of 0.5 is 50\% coarse-grained material (either 100\% sand, 100\% gravel, or some combination of the two) and 50\% fine-grained material (some combination of silt and clay). To assign hydrogeologic parameters (hydraulic conductivity, porosity, etc.) to each coarse fraction realization, we assume that only two types of sediment exist within a single realization --- coarse-grained and fine-grained --- and that each cell is a mix of the two types. We then sample from broad distributions to assign parameter values to each end member and use a weighted averaging procedure to calculate values for cells that are a mix of the two end members. The input parameter distributions are derived from a mix of literature and site-specific data. For more information about these parameter values, we refer the reader to the more complete description provided by \citet{perzan_controls_2023}.

Each MAR simulation consists of three consecutive stages. In Stage 1, a constant recharge forcing equal to the long-term natural recharge rate is applied to the top of the domain for 160,000 days ($\sim$440 years), which is enough time for each simulation to reach steady state. No meteorological or irrigation forcing is applied during Stage 1. In Stage 2, we apply one year of meteorological and irrigation forcing using hourly measurements from the southern Central Valley. These forcings are applied consistently across all MAR simulations. Together, Stage 1 and 2 initialize water content in the vadose zone and establish realistic soil moisture profiles prior to implementing recharge. Finally, Stage 3 simulates recharge by inundating the orchard with $\SI{0.8}{m}$ of water in a single winter season (applied in either 1, 4, 8 or 16 individual recharge events), alternating between periods of flooding and no flooding. In between inundation events, we apply meteorologic and irrigation forcing, which are kept consistent across all MAR simulations, to the top of the domain. Stage 2 begins with the final state of Stage 1, and Stage 3 begins with the final state of Stage 2. In this way, the three stages are dependent.  As discussed by \citep{perzan_controls_2023}, $\sim$53\% of the initial 600 simulations failed to complete all three modeling stages due to invalid parameter combinations. Another 11 simulations successfully reached steady state but lacked time step outputs after reaching steady state; while included in the previous study, they are excluded here to ensure all training examples have consistent time step outputs. Thus, in this study, we examine 273 simulations.

Once each simulation has completed all three stages, we quantify the increase in saturated zone storage induced by MAR. Each simulation produces spatiotemporally variable output (transient pressure fields, represented as a time series of $120 \times 80 \times 25$ rectilinear grids) from a combination of temporally static, spatially variable input (e.g., 3D fields of porosity or saturated hydraulic conductivity, represented as $120\times 80 \times 25$ rectilinear grids) and temporally variable, spatially homogeneous input (e.g., ambient air temperature). The number of output time steps varies depending on the modeling stage, with a finer temporal resolution in later modeling stages. We output pressure fields every 1000 days in Stage 1 (160 output time steps), every 2 days in Stage 2 (183 output time steps) and every 2 days in Stage 3 (366 output time steps). Note that the number of time step outputs here simply reflects the frequency at which the model outputs 3D pressure fields and not the temporal resolution of the underlying solver. From these 3D transient pressure fields, we calculate other system states (e.g., water content in each grid cell) and quantify the change in saturated zone storage.

\subsection{The high-fidelity process-based model: ParFlow-CLM}\label{subsec:parflow}

In the first phase of our proposed framework (\S\ref{subsec:propose}), a process-based hydrologic model is used to generate an initial set of simulations, which are then used to train an ML surrogate model. In this case study, we perform these process-based simulations using ParFlow-CLM. ParFlow-CLM couples the integrated hydrologic code ParFlow with the Common Land Model (CLM) to solve partial differential equations that describe water and energy fluxes over and beneath the land surface. ParFlow solves the Richards' equation for variably saturated subsurface flow using a Newton-Krylov approach and the shallow-wave equations for overland flow using a kinematic wave approximation. CLM, meanwhile, uses a mass transfer approach to compute water and energy fluxes at the land surface \citep{maxwell_development_2005, kollet_capturing_2008}, including irrigation, evapotranspiration, and precipitation. To accurately simulate the impacts of diurnal temperature and radiation fluctuations on water fluxes across the land surface, we run ParFlow-CLM using an hourly time step in Stages 2 and 3 (\S \ref{subsec:casestudy}). For a detailed description of the governing equations, we refer readers to previous work describing ParFlow-CLM \citep{jones_newton-krylov-multigrid_2001, ashby_parallel_1996, kollet_integrated_2006, maxwell_terrain-following_2013, maxwell_development_2005, kollet_capturing_2008}.

\begin{figure}
    \centering
    \begin{subfigure}[t]{0.3\textwidth}
        \centering
        \includegraphics[height=3in]{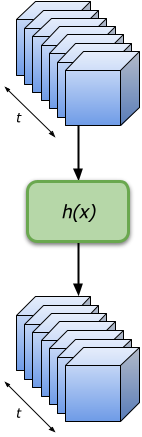}
        \caption{Non-sequential}
        \label{subfig:nonseq}
    \end{subfigure}
    \hfill
    \begin{subfigure}[t]{0.3\textwidth}
        \centering
        \includegraphics[height=3in]{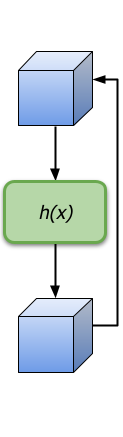}

        \caption{One-step}
        \label{subfig:onestep}
    \end{subfigure}
    \hfill
    \begin{subfigure}[t]{0.3\textwidth}
        \centering
        \includegraphics[height=3in]{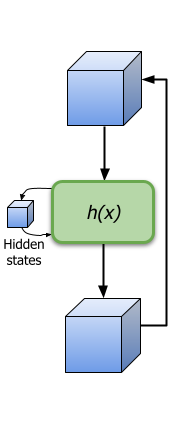}
        \caption{Recurrent}
        \label{subfig:recurr}

    \end{subfigure}

    \vspace{3mm}
    \caption{Illustration of the three types of ML model architectures compared in this study. Each blue cube represents a single 3D tensor. The non-sequential models (a) receive four-dimensional input (\textit{x}, \textit{y}, \textit{z} and time) and produce all pressure fields for all time steps in a single forward pass of the model. The one-step models (b) simulate a single time step per forward pass and roll forward, using the output from previous time steps as input for subsequent steps. The recurrent models (c) use a similar procedure as the one-step models, except that they contain memory cells that retain hidden states between forward passes.}
    \label{fig:model_sum}
\end{figure}

\subsection{The machine learning models}\label{subsec:ml_models}

In the second phase of this hybrid framework (\S\ref{subsec:propose}), we train an ML surrogate on the initial set of simulations generated by the process-based model. In total, we evaluate the performance of seven different machine learning surrogate model architectures: three- and four-dimensional convolutional neural networks, three- and four-dimensional vision transformers \citep{dosovitskiy_image_2021}, three- and four-dimensional U-net Fourier neural operators \citep{wen_u-fnoenhanced_2022,zhou_fedformer_2022,kang_new_2023}, and a three-dimensional recurrent and highway network developed for spatiotemporal forecasting, the PredRNN++ \citep{wang_predrnn_2018}. All seven surrogate architectures are outlined in Table \ref{tab:ml_models}.
We selected these architectures because some of them have been successfully applied in previous hydrologic surrogate modeling studies \citep{tran_development_2021, maxwell_physics-informed_2021,wen_u-fnoenhanced_2022}, while others are some of the most common architectures used for spatiotemporal forecasting in related fields like computer vision.

The ML surrogate architectures can be categorized into three types based on their method of prediction (Table \ref{tab:ml_models}, Figure \ref{fig:model_sum}). First, the \textit{Non-sequential} models emulate the entire simulation in a single forward pass (i.e., the emulator produces 3D pressure fields for all output time steps in one run of the model). By contrast, the \textit{One-step} models emulate one output time step (producing a single 3D pressure field) per forward pass. These models then use the output from the previous time step to perform another forward pass, repeating until all 3D pressure fields have been generated. Finally, the \textit{Recurrent} models are similar to the one-step models in that they emulate one output time step per forward pass. However, they use memory cells to retain hidden states between forward passes, so that information can be stored between output time steps. As discussed above, the output frequency varies between model stages (see \ref{subsec:casestudy}), which results in a different shape input tensor for each model stage. Because most ML architectures require a consistent shape input tensor across all training examples, we train a separate surrogate for each model stage. The ML models are summarized below, though complete details are provided in the Supplementary material (Section \ref{appendix:arch}).

\begin{table}
    \caption{Overview of the principal method, number of parameters, and source for each surrogate model.}
    \label{tab:ml_models}
    \begin{tabular}{c c cc l}
        \toprule
          Model Type & Model Name  &Principal Method& Parameter Count & Source  \\
         \midrule
         &CNN4d  &Convolution& 12,801,281& N/A\\
         (a) Non-sequential & ViT4d  &Attention&18,228,032& \citet{dosovitskiy_image_2021}\\
          & U-FNO4d  &Fourier&251,446,273& \citet{wen_u-fnoenhanced_2022}\\
         \midrule
          & CNN3d  &Convolution& 2,561,601& N/A\\
          (b) One-step & ViT3d  &Attention& 10,238,848 &\citet{dosovitskiy_image_2021}\\
          & U-FNO3d  &Fourier&26,152,993&\citet{wen_u-fnoenhanced_2022}\\
         \midrule
          (c) Recurrent& PredRNN++ &LSTM&11,275,425&\citet{wang_predrnn_2018}\\
          \bottomrule
    \end{tabular}
\end{table}

\paragraph{Non-sequential models}\label{subsubsec:nonseq}
We experiment with three non-sequential models. Since non-sequential models produce 3D pressure fields for all output time steps in a single forward past, they are four-dimensional; i.e., they receive four-dimensional inputs --- \textit{x}, \textit{y}, \textit{z}, and time --- and generate four-dimensional outputs (3D transient pressure fields). The non-sequential models are:
\begin{enumerate}
    \item \textit{CNN4d.} The CNN4d is a three-layer convolutional neural network that consists of four-dimensional convolutional layers, each followed by a four-dimensional batch normalization layer \citep{ioffe_batch_2015} to allow for fast and stable training, a leaky ReLU activation \citep{maas_rectier_2013} to introduce nonlinearities, and a dropout layer \citep{hinton_improving_2012} to prevent overfitting.

    \item \textit{ViT4d.} The ViT4d is a vision transformer \citep{dosovitskiy_image_2021} that uses a four-dimensional convolutional layer to transform the input tensor into a sequence of four-dimensional patches. This sequence of patches is then fed into a self-attention module \citep{vaswani_attention_2017} that calculates the importance of each patch relative to every other patch, allowing the model to learn both spatial and temporal dependencies regardless of their distance from each other.
    \item \textit{U-FNO4d.} The U-FNO4d is based on the three-dimensional U-FNO implementation originally proposed by \citet{wen_u-fnoenhanced_2022}, which consists of a U-shaped network paired with Fourier neural operators \cite[FNOs; ][]{li_fourier_2020}. While a full description of the U-FNO architecture is beyond the scope of this paper, we provide a brief description below. The U-FNO architecture (Figure \ref{fig:arch_ufno}) consists of three Fourier layers followed by three U-Fourier layers, bookended by linear projection layers. Each Fourier layer contains two information pathways: one through a Fourier and inverse Fourier transform separated by a linear transformation and another through a simple linear layer. These two information pathways are merged with a residual connection. The U-Fourier layer is essentially identical to the Fourier layer except there is a third information pathway: a U-Net consisting of a sequence of five convolutional layers followed by three transposed convolutional layers. Again, the information pathways are merged with a residual connection. We expand the original, three-dimensional U-FNO architecture to four dimensions by replacing three-dimensional convolutional layers in the U-Fourier layers with four-dimensional convolutional layers and by adding a fourth dimension with an identical maximum number of nodes to the Fourier and inverse Fourier transforms.
\end{enumerate}

\paragraph{One-step models}\label{subsubsec:onestep}
One-step models, unlike non-sequential models, receive three-dimensional inputs (\textit{x}, \textit{y}, and \textit{z}) and predict one time step at a time. The output of the previous time step is reused as input for the next time step and this rolling forward process continues until the final time step. Note that these models receive no explicit temporal information during training, such as day of year or time step size. These one-step models include:

\begin{enumerate}
    \item[4.] \textit{CNN3d.} The CNN3d model, like the CNN4d model, acts as our baseline one-step model.
    Constructed almost identically to the CNN4d model, the CNN3d model mainly differs in that it employs three-dimensional convolutions instead of four-dimensional ones.
    \item[5.] \textit{ViT3d.} The ViT3d model mirrors the ViT4d model but employs a three-dimensional convolutional layer to generate three-dimensional patches.
    \item[6.] \textit{U-FNO3d.} Our U-FNO3d model replicates the U-FNO architecture presented by \citet{wen_u-fnoenhanced_2022} and is the same as the U-FNO4d described above, except that it uses three-dimensional convolutional layers instead of four-dimensional layers.
\end{enumerate}

\paragraph{Recurrent models}\label{subsubsec:recur}
Like one-step models, one forward pass of a recurrent model produces three-dimensional output for a single time step. The key distinction is that the recurrent model retains hidden states between forward passes. We experiment with one such model:
\begin{enumerate}
    \item[7.] \textit{PredRNN++.} This recurrent model is an improved predictive recurrent neural network (PredRNN++), which was originally developed by \citet{wang_predrnn_2018}. The PredRNN++ architecture (Figure \ref{fig:arch_predrnn}) pairs causal long short-term memory networks (causal LSTMs) with a gradient highway units (GHUs). Causal LSTMs contain a spatiotemporal memory mechanism designed to improve short-term spatial correlations, while GHUs connect future outputs with distant inputs for improved long-term modeling capabilities. Each causal LSTM contains distinct cells designed to store spatial information ($\mathcal M$) and temporal information ($\mathcal C$) from one time step to the next, as well as hidden states ($\mathcal{H}$), which are used to make predictions (Figure \ref{fig:arch_predrnn}). Within the PredRNN++, which contains multiple causal LSTM layers, each layer has its own unique temporal memory ($\mathcal C$) and hidden states ($\mathcal{H}$). The spatial memory ($\mathcal M$) is shared across all layers passed sequentially between them within a single time step. To allow the model to learn long-term dependencies, the GHU layer, equipped with its own memory cell ($\mathcal Z$), is called once per time step, between the first and second causal LSTM layers, and functions similarly to a residual connection. The GHU allows information and gradients to bypass deep-in-time model layers, reducing the risk of vanishing gradients that result from long-term predictions. Given that the original PredRNN++ architecture was designed to handle two-dimensional inputs, we modify the implementation to handle three-dimensional inputs by replacing all two-dimensional convolutional layers with three-dimensional convolutional layers. Detailed descriptions of causal LSTMs and gradient highway unit  can be found elsewhere \citep{wang_predrnn_2018}.
\end{enumerate}

\subsection{Data preprocessing}\label{subsec:data_preproc}

To train each ML surrogate model, we designate the inputs of the ParFlow-CLM simulations (e.g., 3D porosity fields, transient meteorological forcing, etc.) as our training features and the outputs of the ParFlow-CLM simulations (transient, 3D pressure fields) as our training targets. However, prior to training, we first perform several steps of preprocessing. In particular, we perform feature selection to curate a list of salient input features, we broadcast input features across the $x$, $y$, $z$, and time dimensions to standardize input and output formats, and we downsample certain inputs to reduce memory requirements for training.

\newgeometry{margin=2cm}
\begin{landscape}
\begin{spacing}{1.1}
\centering
\begin{threeparttable}
    \caption{Summary of input and target fields. }

    \begin{tabular}{r l c c c c c c c c c}
        \toprule
        &&&&& Varies between & Spatially & Temporally & \multicolumn{3}{c}{Appears in Stage}\\
        \cmidrule{9-11}
        &Field name& Units & $\mu$\tnote{1} & $\sigma$\tnote{1} & simulations & variable & variable & 1 & 2 & 3 \\
        \midrule
        \rowcolor{lightgray}
        &\multicolumn{10}{l}{Inputs}\\
        \midrule
        1. & Horizontal Hydraulic Conductivity & $\unit{m/hr}$ & $0.0936$ & $0.270$
            & \cmark & \cmark & $\xmark$ & \cmark & \cmark & \cmark \\
        2. & Vertical Hydraulic Conductivity & $\unit{m/hr}$ & $0.0151$& $0.0465$
            & \cmark & \cmark & $\xmark$ & \cmark & \cmark & \cmark \\
        3. & Porosity & $1$ & $0.338$ & $0.0585$
            & \cmark & \cmark & $\xmark$ & \cmark & \cmark & \cmark \\
        4. & Residual Saturation & $1$ & $3.417$ & $1.15$
            & \cmark & \cmark & $\xmark$ & \cmark & \cmark & \cmark \\
        5. & van Genuchten $\alpha$ & $\unit{1/m}$ & $3.41$ & $1.15$
            & \cmark & \cmark & $\xmark$ & \cmark & \cmark & \cmark \\
        6. & van Genuchten $n$ & $1$ & $0.365$ & $0.201$
            & \cmark & \cmark & $\xmark$ & \cmark & \cmark & \cmark \\
        \midrule
        7. & Incoming Short-Wave Radiation & $\unit{W/m^2}$ & $202.$ & $86.5$
            & $\xmark$ & $\xmark$ & \cmark & $\xmark$ & \cmark & \cmark \\
        8. & Incoming Long-Wave Radiation & $\unit{W/m^2}$ & $323.$ & $22.3$
            & $\xmark$ & $\xmark$ & \cmark & $\xmark$ & \cmark & \cmark \\
        9. & Precipitation Rate & $\unit{mm/s}$ & $2.85$e-05 & $1.74$e-05
            & $\xmark$ & $\xmark$ & \cmark & $\xmark$ & \cmark & \cmark \\
        10. & Air Temperature & $\unit{K}$ & $290.$ & $6.14$
            & $\xmark$ & $\xmark$ & \cmark & $\xmark$ & \cmark & \cmark \\
        11. & West-to-East Wind Speed & $\unit{m/s}$ & $0.525$ & $0.844$
            & $\xmark$ & $\xmark$ & \cmark & $\xmark$ & \cmark & \cmark \\
        12. & South-to-North Wind Speed & $\unit{m/s}$ & $-0.371$ & $0.902$
            & $\xmark$ & $\xmark$ & \cmark & $\xmark$ & \cmark & \cmark \\
        13. & Barometric Pressure & $\unit{pa}$ & 100,000 & $246.$
            & $\xmark$ & $\xmark$ & \cmark & $\xmark$ & \cmark & \cmark \\
        14. & Specific Humidity & $\unit{kg/kg} $ & $6.40$e-03 & $5.00$e-04
            & $\xmark$ & $\xmark$ & \cmark & $\xmark$ & \cmark & \cmark \\
        \midrule
        15. & Cell Volume & $\unit{m}^3$ & $600.$ & $1130$
            & $\xmark$ & \cmark & $\xmark$ & \cmark & \cmark & \cmark\\
        16. & Initial Pressure Head Field\tnote{2} &$\unit{m}$ & $6.40$ & $18.4$
            & $\xmark$/\cmark & \cmark & $\xmark$ & \cmark & \cmark & \cmark \\
        17. & Inundation Forcing & $\unit{m/s}$ & $0.232$ & $0.294$
            & \cmark & \cmark & \cmark & $\xmark$ & $\xmark$ & \cmark \\
        \midrule
        \rowcolor{lightgray}
        &\multicolumn{10}{l}{Targets}\\
        \midrule
        1. & Pressure Head & $\unit{m}$ & $6.20$ & $21.2$
            & \cmark & \cmark & \cmark & \cmark & \cmark & \cmark\\
        \bottomrule
    \end{tabular}
    \label{tab:sum_fields}
   \begin{tablenotes}
   \begin{footnotesize}
   \item[1] We provide the mean and standard deviation over all 273 simulations. Although not all fields follow a normal distribution, these metrics offer insights into the data's range and variability.
   \item[2] At the start of model Stage 1 (model spin up to steady state), pressure across the domain is in hydrostatic equilibrium with the water table at 45 m depth. For Stages 2 and 3, the initial pressure head field is initialized as the final pressure field output from the previous stage. This reflects the fact that, when these simulations were generated, Stage 2 began simulations with the final state of Stage 1, and, similarly, Stage 3 began simulations with the final state of Stage 2.
   \end{footnotesize}
   \end{tablenotes}
\end{threeparttable}
\end{spacing}
\end{landscape}

\restoregeometry

\subsubsection{Feature selection}

We select 17 input fields as the input features and one output field as our supervised learning target (Table \ref{tab:sum_fields}). These input fields include hydrogeologic parameters (fields 1-6 in Table \ref{tab:sum_fields}), meteorologic forcing (fields 7-14), information about the rectilinear model grid (field 15), initial conditions (field 16), and the recharge forcing applied at the surface (field 17).

The process-based hydrologic model receives additional input information not included in Table \ref{tab:sum_fields}, such as cell dimensions and time step sizes. However, we omit many of these fields from our input features because they are temporally invariant and constant across all training examples. Thus, each surrogate architecture likely learns this information implicitly during training. Some fields, such as cell volume and the initial pressure head for Stage 1 (\S \ref{subsec:casestudy}), are constant across all simulations, but are included nonetheless to support future studies seeking to modify these parameters.

\subsubsection{Broadcasting}

Broadcasting streamlines manipulation of inputs and targets through the use of large, combined tensors. The input features (Table \ref{tab:sum_fields}) include fields that remain constant in time, fields that are constant in space, and fields that are constant in both space and time. Broadcasting simply repeats the values of each field across dimensions where that field remains constant, which ensures that the tensor for each input feature is four-dimensional (\textit{x}, \textit{y}, \textit{z}, and time). For example, we broadcast air temperature at each time step --- which does not vary spatially across the domain --- along the \textit{x}, \textit{y}, and \textit{z} dimensions. Similarly, given that saturated hydraulic conductivity does not change from one time step to the next, we repeat the 3D hydraulic conductivity field from the first time step across all time steps. This approach standardizes the dimensions of all input fields and allows them to be combined into one large tensor for input into each architecture.

\subsubsection{Downsampling}\label{subsubsec:downsampling}

While broadcasting standardizes input dimensions, it increases the total size of the input features and, by extension, the memory allocated to each training example. The memory required to train an ML model is an important component of computational efficiency; if the input and output for a single training example exceeds the memory capacity of a single graphics processing unit (GPU), then the model will have to be trained across multiple GPUs, which will increase the total number of processor-hours used during training. Modern hydrologic model domains can include anywhere from $10^5$ to $>$$10^8$ total grid cells \citep{wood_hyperresolution_2011,bierkens_hyper-resolution_2015}. At this scale, after broadcasting each input feature across the \textit{x}, \textit{y}, \textit{z}, and time dimensions and adding in the size of the training targets (i.e., the transient 3D pressure fields from ParFlow-CLM), the memory allocated to a single training example can quickly exceed 100 GB, which is larger than the memory capacity of most modern GPUs. To address this challenge, we limit the size of a single training example through temporal and spatial downsampling.

For Stages 1 and 2 (\S \ref{subsec:casestudy}), we apply temporal downsampling. These model stages are only used to initialize water content throughout the vadose zone. Because intermediate pressure fields are not used to calculate recharge efficiency, decreasing the temporal resolution does not limit downstream analyses. For each stage, we sample 16 output time steps from the spin up period using a geometric sampling scheme:

\begin{equation}
    \label{eqn:geometricsampling}
    \left\{\left\lfloor \sum_{k=1}^t r^{k-1}\right\rfloor : \sum_{k=1}^{16} r^{k-1} = 160 \right\}_{t=1}^{16}
\end{equation}

The length of 16 is chosen so that a single training example fits on our training hardware, while the geometric sampling scheme prioritizes time step outputs toward the beginning of the model stage, when changes in pressure head are most pronounced. This flexibility in time step outputs emphasizes the advantage of generating simulation outputs with ML models: while a process-based model may require short time steps to converge to a solution, an ML model can reasonably learn patterns across larger time scales and can take arbitrarily large time steps. Though the final steady-state pressure field at the end of each stage is the only output used in subsequent modeling stages, we calculate loss across multiple time steps from this stage to ensure that the model simulates physically meaningful processes (i.e., gradual changes in pressure over time). This also forces the model to output intermediate pressure fields; visually examining these intermediate outputs allows the user to more easily diagnose factors contributing to poor model performance. While this geometric sampling scheme results in a variable time step size within an individual simulation, the interval between sampled time steps is consistent across all training examples. Thus, the data-driven ML models should implicitly learn these temporal patterns during training.

To accurately quantify recharge efficiency, the output from the Stage 3 surrogate models must be temporally high-resolution. Thus, instead of applying temporal downsampling, we apply spatial downsampling to the Stage 3 input features and training targets. For the input features, we reduce dimensionality using an average pooling layer with a $20\times 20\times 1$ window and a stride of $4\times 4\times 1$. This layer simply calculates the mean of each input feature over the $20\times 20\times 1$ window, reducing the size of the input tensor in the $x$ and $y$ directions. The $z$ dimension remains unchanged, however, in order to preserve vertical resolution given that vertical flow is important for this case study.

\begin{figure}
    \centering
    \begin{subfigure}[t]{\textwidth}
        \centering
        \includegraphics[width=6in]{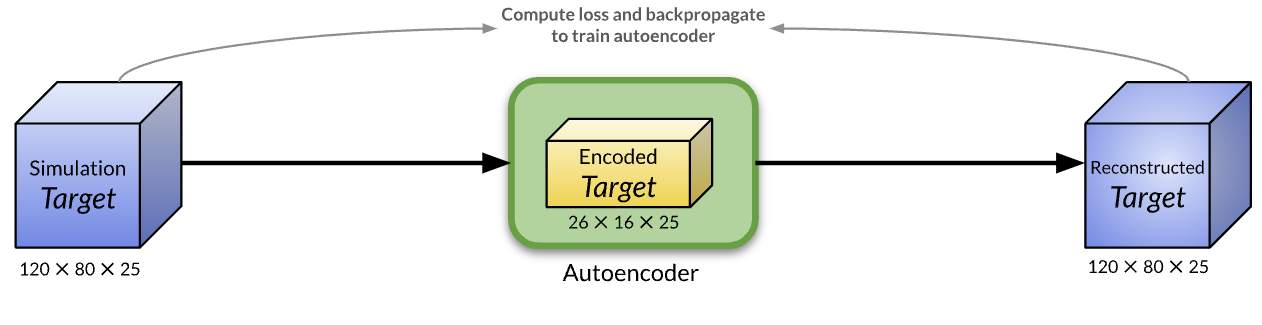}
        \caption{Autoencoder training procedure.}
        \label{fig:autoenc_a}
    \end{subfigure}

    \vspace{20mm}

    \begin{subfigure}[t]{\textwidth}
        \centering
        \includegraphics[width=7in]{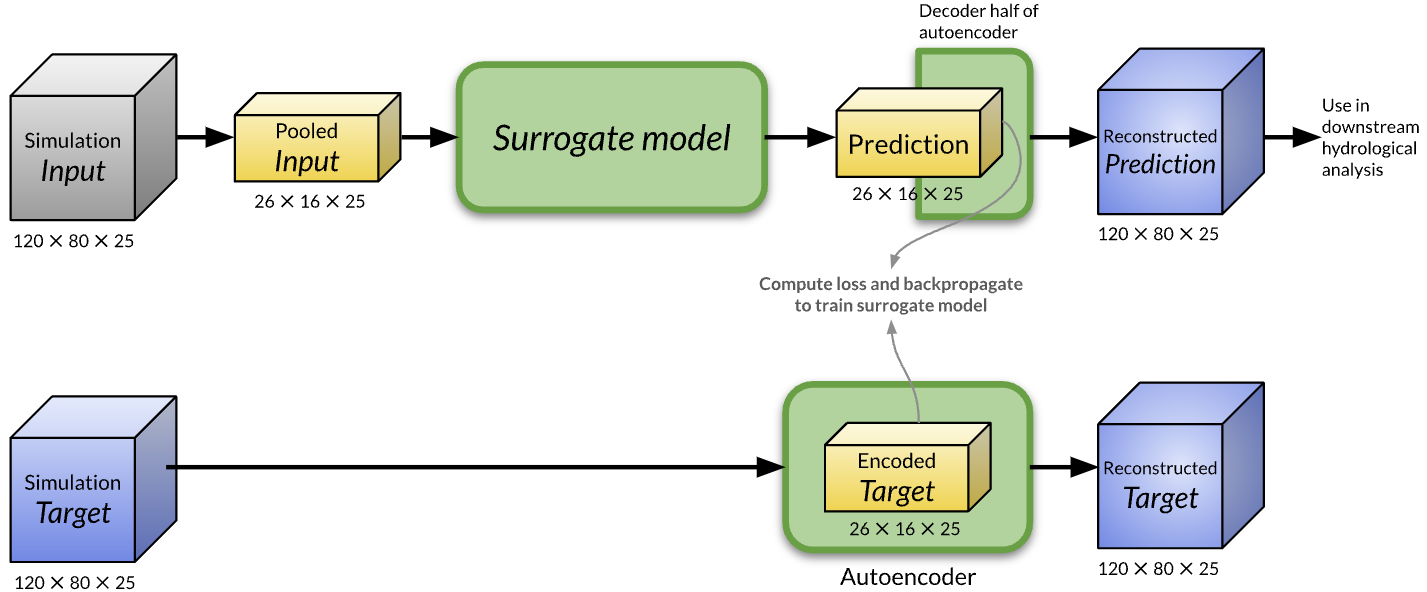}
        \caption{Procedure to train a surrogate model in a reduced dimensional space. Autoencoder weights are fixed.}
        \label{fig:autoenc_b}
    \end{subfigure}
    \caption{Workflows to train an autoencoder (a) and subsequently use the autoencoder in the surrogate model's training (b).}
    \label{fig:autoenc}
\end{figure}

To downsample the Stage 3 training targets, we use an autoencoder. The autoencoder is a separate neural network that is trained to reproduce the spatially variable pressure field from a single time step, but it is forced to store this information in an intermediate, lower-dimensional space (Figure \ref{fig:autoenc_a}). Each surrogate model is then trained on the lower-dimensional representation of the 3D pressure fields, known as the encoded targets (Figure \ref{fig:autoenc_b}). We train a single autoencoder that is used for all surrogate models. To calculate surrogate performance on the test set, we then decode the surrogate model's predictions using the autoencoder and calculate error metrics relative to the high-resolution pressure fields output by ParFlow-CLM. This workflow allows us to train our surrogate model in the reduced dimensional space, where tensors are 4\% their original size, while retaining the high spatial resolution of our predictions for use in downstream tasks, such as calculating recharge efficiency. For a detailed overview of the autoencoder architecture, please refer to Section \ref{appendix:autoenc} in the Supplementary material.

\subsection{Model training and evaluation}\label{subsec:exp_setup}

In this section, we discuss the training setup used for all surrogate models, the hardware used to calculate computational efficiency and the evaluation criteria used to quantify model performance.

\subsubsection{Training configuration} \label{subsubsec:config} Each model undergoes training for $150$ epochs with a learning rate of $0.0001$, using Adam optimization \citep{kingma_adam_2015}. Due to GPU memory constraints, we restrict the batch size to $1$. We employ a normalized L2 loss function:

\begin{equation}
    \label{eqn:normalizedloss}
    L({y}_i, h({x}_i))= \frac{||w \cdot ({y}_i - h({x}_i))||_2}{||{y}_i||_2}
\end{equation}

where $y_i$ is the ground-truth output for the $i$th training example, $x_i$ is the input for the $i$th training example, $h(x_i)$ is the predicted output, $w$ defines a depth-wise weighting scheme, and  $||\cdot||_2$ is the L2 norm. We choose this normalized loss function because it is particularly effective when there is large variance in the data and because previous hydrologic surrogate modeling work suggests that it provides faster convergence than the L1 loss \citep{wen_u-fnoenhanced_2022}. Given that managed aquifer recharge efficiency is largely governed by unsaturated flow processes \citep{perzan_controls_2023}, accurately simulating pressure changes within the vadose zone is more important than simulating pressure changes deep within the saturated zone. Thus, we define the weights $w$ such that they more severely penalize pressure errors closer to the surface than at depth (Figure \ref{fig:relative_weights}).

To investigate the impact of our choice of loss function on model performance, we train additional CNN3d surrogates on Stage 1 (\S \ref{subsec:casestudy}) simulations using the mean square error loss (i.e., the square of the numerator of Equation \ref{eqn:normalizedloss}). Due to computational constraints, we could not train duplicates of each surrogate architecture using the mean square error loss, so we only perform this comparison with the CNN3d because it has the fastest training time.

We perform a random 80\%-10\%-10\% (217-28-28) training--validation--test split and all performance metrics are reported based on the test set unless otherwise noted. Features and labels are normalized using their respective minimum and maximum values from the training set. We normalize all splits by the training set's minimums and maximums to ensure the model does not inadvertently gain knowledge of the validation and test sets before it is supposed to evaluate on them. To assess the effects of data normalization on ML model accuracy, we train two additional Stage 1 CNN3d models: one using Z-score normalization and one without any data normalization. As discussed above, we only perform these tests using the CNN3d because of its short training time.

\subsubsection{Hardware environment} \label{subsubsec:hardware} We train all ML models on a single NVIDIA A100 Tensor Core graphics processing unit (GPU) with 80 GB memory. All ParFlow-CLM simulations were performed on AMD Epyc 7543 central processing units (CPUs). Depending on the model stage (\S \ref{subsec:casestudy}), ParFlow-CLM simulations were run either on 4 MPI ranks with 1 CPU core per rank (Stage 1 and Stage 3) or 6 MPI ranks with 1 CPU core per rank (Stage 2).

\subsubsection{Model accuracy evaluation} \label{subsubsec:e2e}
We evaluate each surrogate model using the mean absolute percentage error (MAPE) between the predicted pressure head time-series and the ground-truth (ParFlow-CLM) pressure head time-series in a cell-by-cell manner. Because we train a different surrogate for each modeling stage, we report different MAPE values for each stage. As with the loss function, we weight MAPE by depth to ensure that each model's performance accurately reflects its ability to simulate vadose zone flow (Figure \ref{fig:relative_weights}). We also evaluate model performance when each surrogate is run sequentially across all three stages (\S \ref{subsec:casestudy}), in what we call an end-to-end (E2E) evaluation. The only difference between a simulation generated during training and a simulation generated in an E2E manner is that, during training, the initial pressure head fields (field 16 in Table \ref{tab:sum_fields}) of Stages 2 and 3 are populated with the ground-truth initial pressure head field produced by ParFlow-CLM. During E2E evaluation, the initial pressure head fields of Stages 2 and 3 are populated with the final pressure head output of Stages 1 and 2, respectively. E2E provides a more realistic evaluation of surrogate performance in a real-world application.

Training an accurate ML surrogate requires a sufficiently large training set, but at a certain point, the cost of increasing the training set size may outweigh any benefit in improved prediction performance. To evaluate this relationship and estimate the minimum number of process-based simulations required to train an accurate surrogate, we train multiple versions of each Stage 1 surrogate model while using smaller and smaller training sets. We only consider models trained on Stage 1 simulations in this section to simplify analysis and minimize additional training. In addition to the baseline training set size of 245 examples used for training (with a 217-28 training-validation split), we use training set sizes of 196 (174-22 split), 147 (130-17), 98 (87-11), 49 (43-6), 25 (22-3), 13 (11-2), 8 (7-1), 5 (4-1), and 3 (2-1) training examples while maintaining the test set constant at 28 examples. Because ML models typically exhibit a power-law relationship between model error and training set size \citep{hestness_deep_2017}, we then fit a power function to these data ($y=am^b$, where $m$ is the number of training examples, $y$ is the test set MAPE, and $a$ and $b$ are regression coefficients). Lastly, we differentiate $am^b$ with respect to $m$, which yields $abm^{b-1}$. This expression represents the instantaneous change in MAPE given an increase in training set size, which we use to compare each model's sensitivity to training set size.

\subsubsection{Model runtime evaluation}\label{subsubsec:runtime_eval}
In addition to evaluating ML model accuracy, we quantify the total compute time needed to generate simulations with this hybrid modeling workflow. Since model runtime is largely dependent on the hardware on which it is executed, we strictly adhere to the hardware environment described in \S\ref{subsubsec:hardware} when evaluating model runtime. To simplify analysis and minimize additional training, we only consider models trained on Stage 1 (\S \ref{subsec:casestudy}) simulations in this section.

First, we benchmark runtime by defining the time needed to generate $N$ Stage 1 simulations using the processed-based hydrologic model, ParFlow-CLM, as
\begin{equation}
    \label{eqn:t_pf}
    T_\text{PF} = N t_\text{PF},
\end{equation}
where $t_\text{PF}$ denotes the CPU-hours taken by ParFlow-CLM to generate a single Stage 1 simulation (i.e., the wall clock time multiplied by the number of CPU cores used for the parallel simulation). Though runtime varies across stochastic ParFlow-CLM simulations (1.07--19.37 CPU-hours for Stage 1 simulations), we calculate $t_\text{PF}$ here as the mean runtime across all Stage 1 simulations (4.17 CPU-hours).

We also establish the number of GPU-hours required to output $N$ Stage 1 simulations using an ML model as
\begin{equation}
    \label{eqn:t_ml}
    T_\text{ML} = T_\text{train} + (N - \mathcal D_\text{train}) t_\text{ML},
\end{equation}
where $T_\text{train}$ denotes the training time needed to reach the minimum MAPE on the validation set, $t_\text{ML}$ denotes the time taken by the trained ML model to generate one Stage 1 simulation, and $\mathcal D_\text{train}$ represents the minimum training set size necessary for the ML model to achieve an MAPE of less than $10\%$ --- the threshold for ``highly accurate forecasting'' \citep{lewis_industrial_1982}. Due to differences in each surrogate model architecture, the minimum training set size ($\mathcal D_\text{train}$) differs for each model. To find this value, we train multiple versions of each surrogate model while varying training set sizes using a binary search approach. The test set (28 simulations) is held constant during this procedure and is used to evaluate models, ensuring they fall below the $10\%$ MAPE threshold. We then calculate $\mathcal D_\text{train}$ as the smallest training set that still allows for a test set MAPE below 10\%. Note that $\mathcal D_\text{train}$ includes both training and validation examples with an 80\%-10\% train--validation split, so each model trains on $\frac 89 \cdot \mathcal D_\text{train}$ examples and validates using $\frac 19 \cdot \mathcal D_\text{train}$ examples.

Given that the training set is generated by a process-based model, the total compute time used to generate $N$ simulations in this hybrid workflow includes both Equation \ref{eqn:t_ml} and the time to generate the training set:
\begin{equation}
    \label{eqn:trainingset}
    T_\text{training set} = \mathcal D_\text{train} \cdot t_\text{PF}.
\end{equation}

Equations \ref{eqn:t_pf}, \ref{eqn:t_ml}, and \ref{eqn:trainingset} have different units (CPU-, GPU-, and CPU-hours, respectively) because the ML model and the process-based hydrologic model run on different types of processors. Depending on the computing resources available to an individual modeler, these units might not be equivalent. Nonetheless, it is  valuable to compare these compute times in order to evaluate the efficiency of a given surrogate model against a traditional process-based model. By assuming that $1 \text{GPU-hour} = 1 \text{CPU-hour}$, we quantify the compute time saved using this hybrid modeling workflow as
\begin{align}
    T_\text{saved} &= N t_\text{PF} - \Big(T_\text{training set} + T_\text{train} + (N - |\mathcal D_\text{train}|) t_\text{ML}\Big)\nonumber \\
    &= \underbrace{(t_\text{PF} - t_\text{ML})}_\text{slope} N - \underbrace{(|\mathcal D_\text{train}| (t_\text{PF} - t_\text{ML}) + T_\text{train})}_\text{intercept},
    \label{eq:saved_runtime}
\end{align}

where $T_\text{saved}$ is in generic processor-hours. The slope term in Equation \ref{eq:saved_runtime} indicates that the time saved by employing an ML surrogate model becomes more pronounced when $t_\text{ML}$ is significantly lower than $t_\text{PF}$. Furthermore, the intercept term indicates that time savings begin sooner, i.e., at smaller values of $N$, if the required training set is small and the training time is low, as expected.


\subsection{Summary of experiments}\label{subsec:experiments}
In summary, we trained six different groups of surrogate models, each for a different numerical experiment:
\begin{enumerate}
    \item We train all seven surrogate architectures --- the CNN4d, ViT4d, U-FNO4d, CNN3d, ViT3d, U-FNO3d, and PredRNN++ (\S\ref{subsec:ml_models}) --- on Stage 1, 2, and 3 ParFlow-CLM simulations (\S \ref{subsec:casestudy}), amounting to 21 trained models. We also train an autoencoder for all Stage 3 models. This totals 22 trained models.
    \item We train two additional Stage 1 CNN3d models using alternative data normalization techniques (\S \ref{subsubsec:config}). One of the additional models does not use any data normalization, and the other uses Z-score normalization. The Stage 1 CNN3d model is chosen over other stages and models for its efficiency in training and evaluation.
    \item We train six additional Stage 1 CNN3d models using different combinations of loss function and data normalization (\S \ref{subsubsec:config}). Three CNN3d models use the normalized L2 loss with min-max, Z-score, and no data normalization, and three CNN3d models use the mean squared error loss with min-max, Z-score and no data normalization.
    \item To evaluate the effect of training set size on model accuracy, we train all seven ML models with nine Stage 1 training sets of decreasing sizes, as described in \S\ref{subsubsec:e2e}. This experiment includes 63 trained models in total.
    \item To evaluate the ML models' runtime efficiency, we re-train all seven ML models with variable Stage 1 training set sizes, using a binary search technique to find the minimum training set size that achieves $<$10\% test MAPE. The number of models trained at this stage varies according to the efficiency of the search algorithm and the test MAPE results from previous experiments but ranged from 2 to 7 across the seven architectures. We then compare the runtimes of the seven models trained on their respective minimum Stage 1 training sets with the procedure described in \S\ref{subsubsec:runtime_eval}.
\end{enumerate}

\section{Results and Discussion}

Balancing speed and accuracy, the hybrid modeling framework achieves runtime benefits with minimal accuracy trade-offs. In \S\ref{subsec:model_acc}, we first present the predictive accuracy of each ML model when trained on the full training set before examining the impacts of training set size and normalization on model accuracy and training stability. In \S\ref{subsec:runtime_analysis}, we compare the computational efficiency of each ML surrogate with a baseline approach that relies solely on the process-based hydrologic model. Lastly, we discuss the importance and challenges of model interpretability (\S\ref{subsec:interpret}) and potential limitations of this framework (\S\ref{subsec:limits}).

\subsection{Model accuracy}\label{subsec:model_acc}

\begin{figure}
    \centering
    \includegraphics[width=0.6\textwidth]{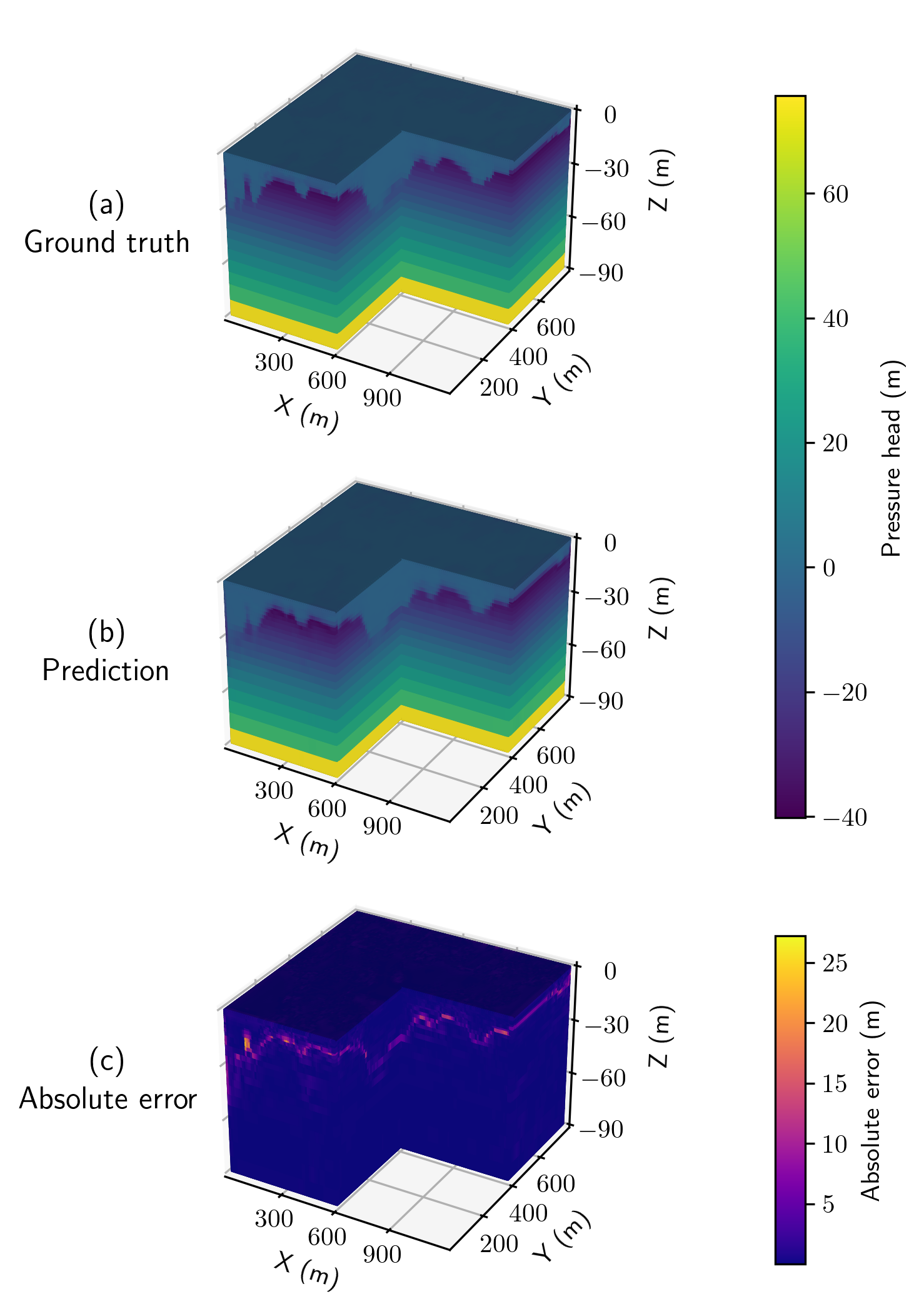}
    \caption{Snapshot of the 3D pressure field for a random test set example at a single time step in Stage 1, as produced by the process-based hydrologic model (a) and the PredRNN++ (b). The cell-by-cell absolute error (c) between the ground truth and predictions is greatest near the wetting front. Note that we omit the deepest layer, which has a height of 60 $\si{m}$, for visual clarity.}
    \label{fig:one_ex}
\end{figure}

Each ML surrogate model produces pressure fields that are visually similar to those produced by the process-based hydrologic model. For example, a comparison of ParFlow and the PredRNN++ shows that both models produce similarly-shaped heterogeneous wetting fronts during Stage 1 (Figure \ref{fig:one_ex}), with monotonically increasing pressure values below the front. Plotting the cell-by-cell absolute error between the two models reveals that differences are greatest near the wetting front (Figure \ref{fig:one_ex}c). However, absolute pressure differences are generally less than a few meters, which is of a reasonable range given the sharp hydraulic gradients that arise near infiltration wetting fronts.

\begin{figure}
    \centering
    \includegraphics[width=5in]{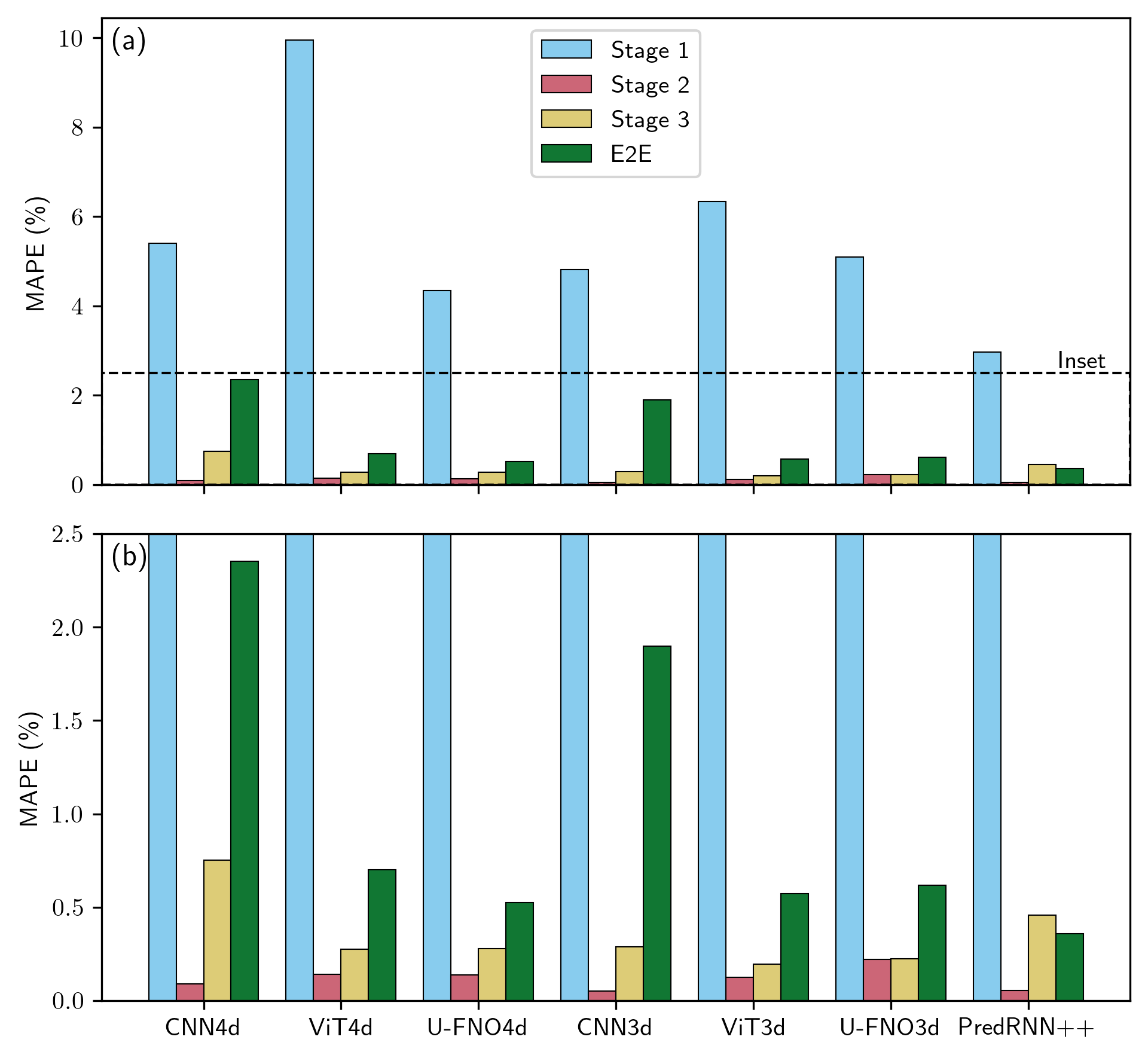}
    \caption{Test set MAPEs for all ML models across all stages. (b) is a local enlargement of (a), indicated by the ``Inset'' label in (a). The ML surrogate models are grouped by model type (see \S \ref{subsec:ml_models}), with the non-sequential models on the left (CNN4d, ViT4d and U-FNO4d), the one-step models in the middle (CNN3d, ViT3d and U-FNO3d) and the recurrent model on the right (PredRNN++). Exact MAPE values are listed in Table \ref{tab:res_compile}.}
    \label{fig:mape_all_stages}
\end{figure}

All seven ML models achieve $<$10\% MAPE on the test set (Figure \ref{fig:mape_all_stages} and Table \ref{tab:res_compile}), which is the threshold for ``highly accurate forecasting'' \citep{lewis_industrial_1982}. The PredRNN++ displays the lowest error for Stage 1 and end-to-end (E2E) predictions, while the CNN3d and ViT3d perform the best on Stage 2 and Stage 3, respectively. Even though the non-sequential models have more parameters than their one-step counterparts (Table \ref{tab:ml_models}), they do not exhibit an appreciable increase in predictive performance. This result suggests that model size alone is not a reliable indicator for model performance and the type of architecture (one-step vs. non-sequential) may have a stronger impact on model performance. One architectural property that lends an advantage for spatiotemporal predictions is the use of large memory cells in a recurrent manner by the PredRNN++. Indeed, the PredRNN++ outperforms the others in E2E evaluation, the measurement most representative of real-world applications. The large memory cells retained between time steps likely provide ample degrees of freedom for the model to store spatially heterogeneous information over time, and these information stores are referenced recurrently, honestly reflecting the step-by-step manner in which the simulations are generated by the ground-truth ParFlow-CLM.

Overall, prediction error is the highest for Stage 1, with a mean MAPE of 5.6\% across all ML models, as opposed to 0.13\% and 0.42\% for Stages 2 and 3, respectively. The higher Stage 1 MAPE could result from its large time scale (\S\ref{subsec:casestudy} and \ref{subsubsec:downsampling}). While Stages 2 and 3 span 1 and 2 years, respectively, Stage 1 spans over 400 years, with more dramatic changes in pressure between outputs. Pressure head changes across large time steps may be more difficult to learn than smaller, more incremental changes, potentially contributing to the poorer performance of Stage 1 models. Sampling the spin-up period at higher temporal resolution (\S \ref{subsubsec:downsampling}) may improve surrogate model accuracy.

The higher MAPE in Stage 1 could also be explained by the fact that its pressure head field is initialized not with a ground-truth pressure head field but with pressure values in hydrostatic equilibrium with the water table at 45 m depth (Table \ref{tab:sum_fields}). By contrast, the pressure heads of Stages 2 and 3 are initialized with the ground-truth pressure head field output by ParFlow-CLM at the end of Stages 1 and 2, respectively. Previous modeling work at this site has shown that managed aquifer recharge efficiency is most strongly controlled by vadose zone water content prior to inundation \citep{perzan_controls_2023}. Thus, stages that use ground-truth initial pressure head fields as inputs (Stages 2 and 3) achieve superior performance than the stage that does not (Stage 1).

\begin{figure}
    \centering
    \includegraphics[width=\linewidth]{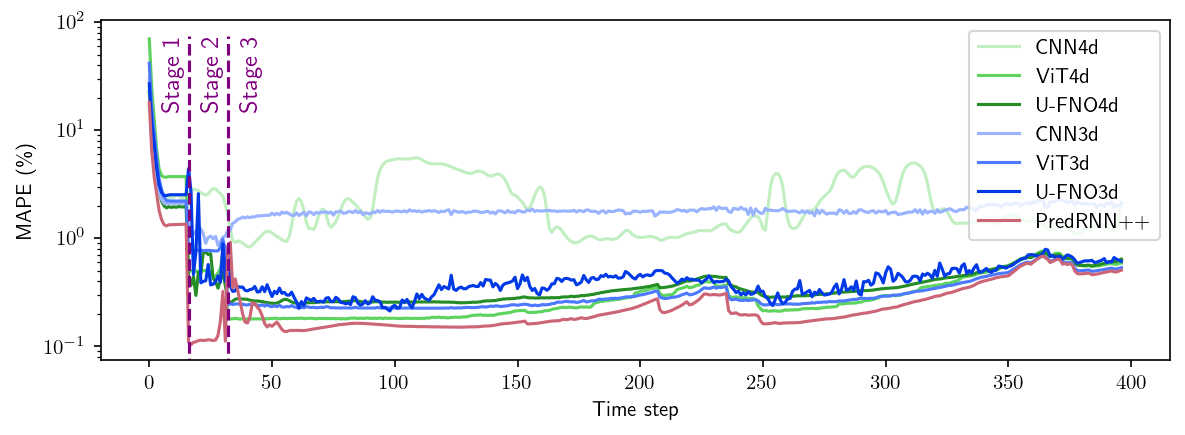}
    \caption{Mean absolute percentage error (MAPE) per time step for all ML models across all model stages. Not all time steps are equally sized; Stage 1 (16 time steps), Stage 2 (16 time steps) and Stage 3 (366 time steps) each encompass 440 years, 1 year, and 2 years of simulation time, respectively. The $y$-axis, MAPE (\%) has a log scale to emphasize minute differences in later time steps. Note that calculating the average MAPE over the first 16 time steps corresponds to the Stage 1 MAPEs presented in Figure \ref{fig:mape_all_stages}; this is not necessarily true for subsequent stages due to the E2E evaluation method.}
    \label{fig:e2e}
\end{figure}

End-to-end (E2E) predictions, on the other hand, do not receive ground-truth pressure fields between stages; the predicted pressure at the end of one stage is used as input for the ML surrogate in the next stage. Overall, the MAPE for E2E predictions is typically lower than that of Stage 1 and similar in magnitude to that of Stage 3 (Figure \ref{fig:mape_all_stages}). The similarity between Stage 3 and E2E error likely arises because E2E error is averaged equally over the entire time series and Stage 3 has the most time steps. However, not all models follow this trend; the CNN3d and CNN4d both exhibit E2E MAPE higher than 1\%, with a $3.1\times$ and $6.6 \times$ respective increases compared to their Stage 3 MAPEs. Plotting the E2E MAPE per time step in Figure \ref{fig:e2e} reveals that the CNN3d and CNN4d exhibit the highest MAPE per time step among other models in Stage 3. This heightened difference between Stage 3 MAPE and Stage 3 MAPE in E2E evaluation could indicate that the CNN3d and CNN4d both exhibit a strong dependence on the initial pressure head at the start of a model stage. The curves for all other models follow the same general shape, suggesting that these models may have learned similar patterns in the Stage 3 data that have resulted in similar accuracy.

Collectively, the results in this section suggest that ML surrogate models can simulate variably saturated groundwater flow with a similar level of accuracy as process-based hydrologic models, supporting the feasibility of applying a hybrid modeling framework to this case study.

\subsubsection{Impact of training set size on accuracy}

\begin{figure}
    \begin{subfigure}{0.49\textwidth}
        \centering
        \includegraphics[width=3.25in]{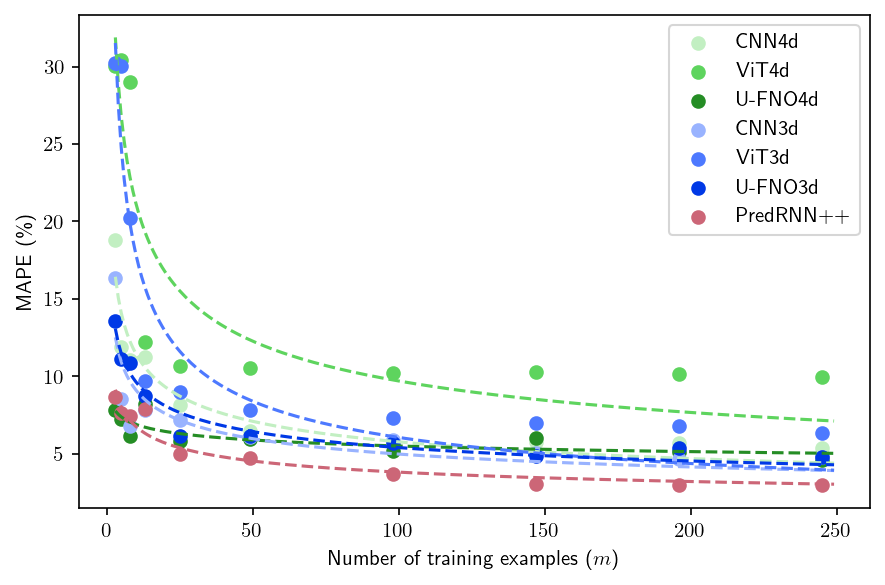}
    \end{subfigure}
    \hfill
    \begin{subfigure}{0.49\textwidth}
        \centering
        \includegraphics[width=3.25in]{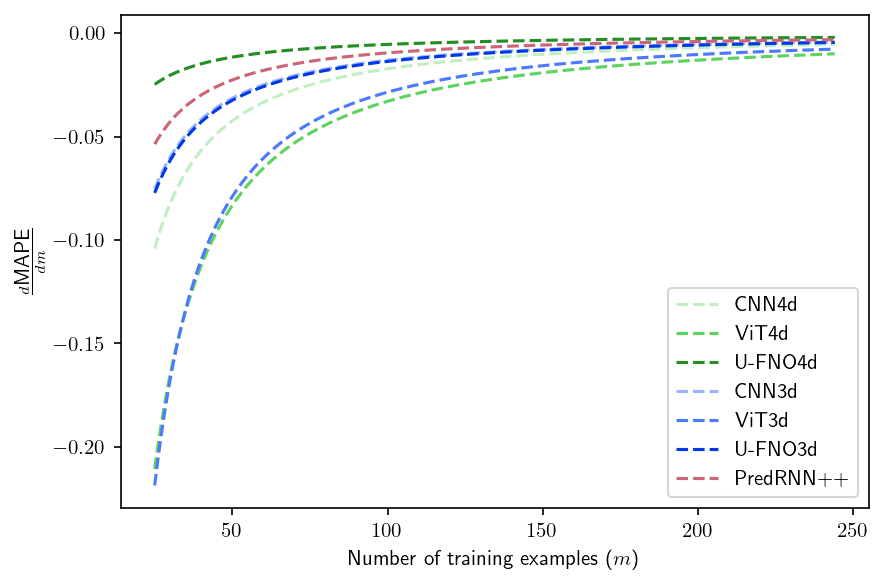}
    \end{subfigure}
    \caption{
    The mean absolute percentage error (MAPE) of the predicted pressure head achieved by models trained on training sets of various sizes (a) and the change in MAPE per unit increase in the training set size (b). The points in (a) represent the MAPE of each trained model, while the dashed lines correspond to a best-fit power law relationship. The lines in (b) are then calculated as the slope of this power law relationship.}
    \label{fig:train_size_mape}
\end{figure}

While the PredRNN++ exhibits the lowest E2E error, it also demonstrates the second lowest sensitivity to training set size. Figure \ref{fig:train_size_mape} reveals that the ML models, in general, exhibit a power-law relationship between training set size and MAPE, which is typical for most deep learning architectures \citep{hestness_deep_2017}. This power-law relationship demonstrates diminishing returns when increasing training set size beyond a certain point. To explore these diminishing returns more closely, we plot the derivative of the best-fit curve for this power-law relationship ($\frac {d\text{MAPE}}{dm}(m)$) in Figure \ref{fig:train_size_mape}. The $\frac {d\text{MAPE}}{dm}(m)$ values for the PredRNN++ remain consistently close to 0, indicating a low sensitivity to changes in training set size, and the model is able to achieve $\sim$5\% MAPE with as few as 25 training examples (Figure \ref{fig:train_size_mape}). In fact, almost all models exhibit nearly consistent MAPE values after 50--100 training examples. While the models presented in Figure \ref{fig:mape_all_stages} were trained on 245 examples (with a 217-28 training-validation split), these results suggests that future practitioners could achieve similar performance by training on fewer examples. In \S\ref{subsec:runtime_analysis}, we even find that some hydrologic surrogate models such as the U-FNO4d and PredRNN++ can achieve $<$10\% MAPE with as low as 3 training examples. However, the relationship between training set size and MAPE is highly dependent on the specific problem of interest, so future practitioners should use caution when extending these results to other sites or other scenarios.

\subsubsection{Impact of normalization on accuracy}\label{subsec:ablat}

\begin{figure}
    \centering
    \begin{subfigure}{\textwidth}
        \centering
        \includegraphics[width=4in]{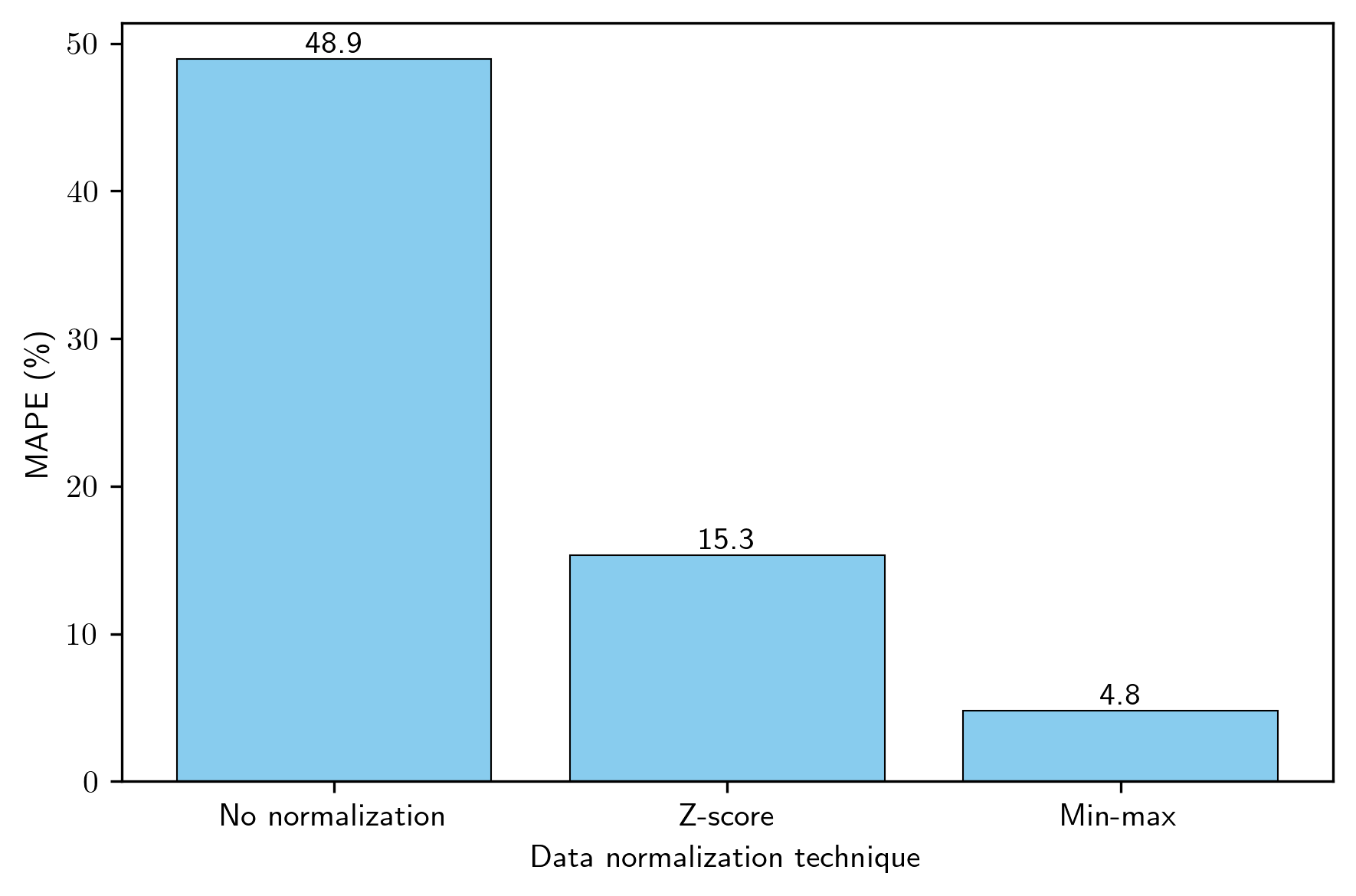}
    \end{subfigure}
    \begin{subfigure}{\textwidth}
        \centering
        \includegraphics[width=4in]{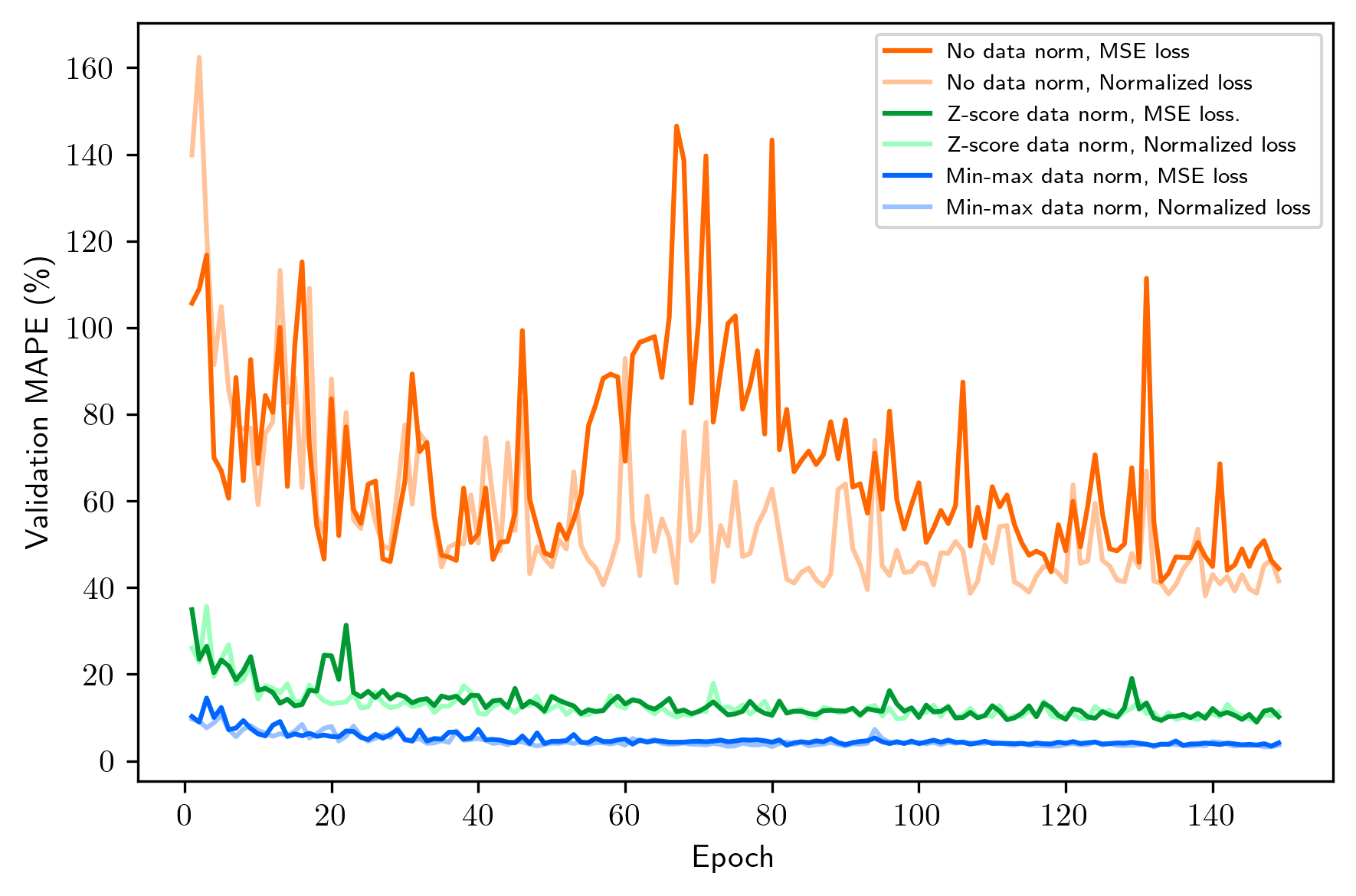}
    \end{subfigure}
    \caption{
    Test set MAPE for Stage 1 CNN3d models trained on various data normalization conditions (a) and test set MAPE at each training epoch for Stage 1 CNN3d models trained using different loss functions (b). Min-max data normalization helps achieve lower MAPE, and normalized loss helps achieve more stable training.}
    \label{fig:norm}
\end{figure}

Both the min-max data normalization and the normalized loss function (\S\ref{subsubsec:config}) improve model performance, resulting in lower MAPE and a more stable training process. When trained on a Stage 1 dataset normalized with a min-max procedure, the CNN3d achieves 4.8\% MAPE, which is an order of magnitude lower than the CNN3d trained without any normalization (48.9\% MAPE) and three times lower than the model trained with a Z-score normalized dataset (15.3\% MAPE; Figure \ref{fig:norm}).  We attribute this improved performance to the variability in the input dataset (Table \ref{tab:sum_fields}), which contains several features that span multiple orders of magnitude. In such cases, normalization improves accuracy by constraining features to a consistent range. In addition, some of the input features are not necessarily normally distributed. Because min-max normalization preserves distribution information --- as opposed to Z-score normalization, which standardizes each input feature to have a mean of 0 and a standard deviation of 1 --- it provides this distribution information to each ML model, which may improve model accuracy.

While normalizing input data improves model accuracy, loss normalization increases training stability. Comparing MAPE on the validation set across all training epochs, we see that using a normalized loss generally reduces the number of sharp spikes in MAPE, indicating that a normalized loss allows for more stable descent and reliable training convergence. One reason loss normalization may have such a large impact in this case is our use of small batch sizes. To prevent memory overflow in the PredRNN++, we train with a batch size of 1; we use a batch size of 1 for all other models to maintain consistent training configurations. However, at small batch sizes, the gradient is calculated over fewer examples (or in this case, a single example), which leads to noisy loss curves and increases the likelihood that gradient descent can get stuck in a local minimum. Loss normalization helps offset these drawbacks by ensuring that the gradients are of consistent magnitude across the training set. This effect is most noticeable in models that use either no data normalization or Z-score data normalization, suggesting that loss normalization most benefits models with inadequate data normalization.

Due to computational constraints, we only perform these analyses with the CNN3d. However, we expect that results would be similar or even more pronounced for other architectures; the CNN3d has built-in batch normalization layers that increase training stability, while most other architectures do not. Overall, these results underscore the importance of both data normalization and loss normalization in enhancing model performance and training stability for hydrologic surrogate models.

\subsection{Computational efficiency}\label{subsec:runtime_analysis}

\begin{figure}
    \centering
    \includegraphics[width=4in]{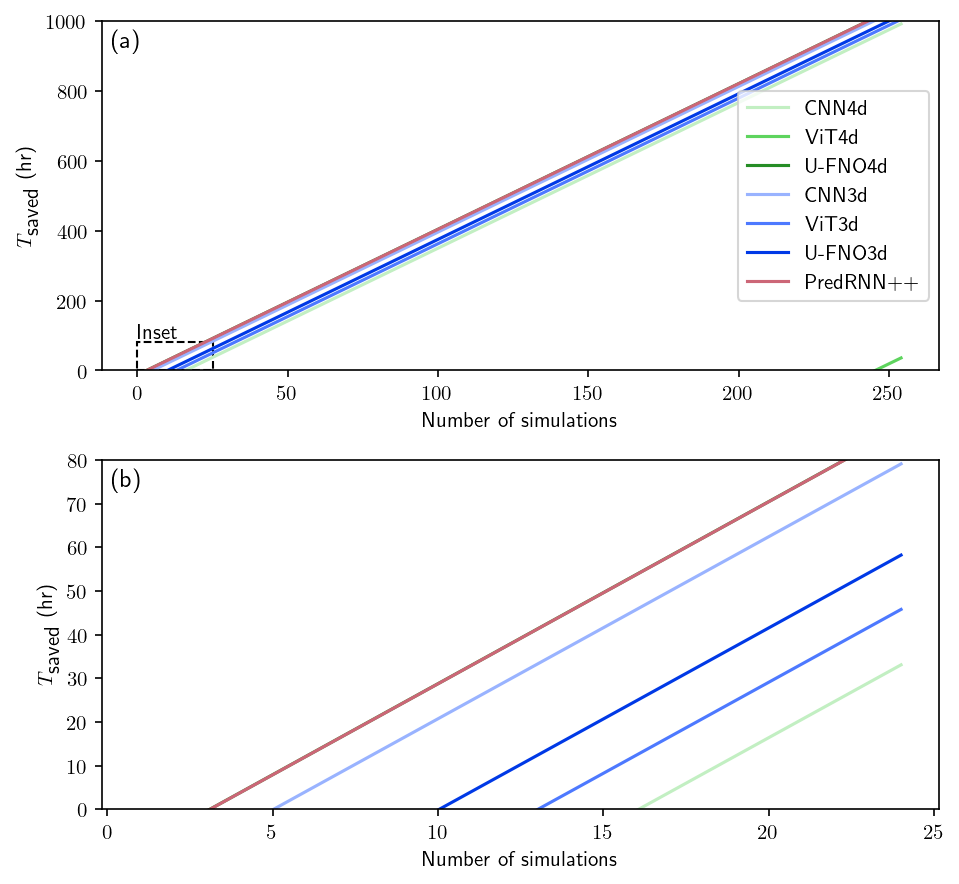}
    \caption{The time saved, in processor-hours, when performing a given number of simulations using an ML model as compared to using ParFlow-CLM. (b) is a local enlargement of (a), with the enlarged region labeled with ``Inset'' in (a). $T_\text{saved}$ is calculated following Equation \ref{eq:saved_runtime} and includes both training time and evaluation time. Note that the U-FNO4d line is almost entirely overlapped by the PredRNN++ line.}
    \label{fig:runtime_inset}
\end{figure}

While each ML surrogate model achieves $<$$10\%$ MAPE, the models vary widely in the amount of time and memory they require to produce simulations. One attribute they share, however, is that all ML surrogate models offer significant runtime advantages over the process-based hydrologic model (Figure \ref{fig:runtime_inset}). Figure \ref{fig:runtime_inset} reveals positive slopes across all lines. Recall that the slope for these plots are calculated as the difference in time taken by ParFlow-CLM and the ML model in generating a single Stage 1 simulation (Equation \ref{eq:saved_runtime}). All lines appear to exhibit identical slopes because each ML surrogate outputs simulations orders of magnitude faster than the process-based hydrologic model ($t_\text{PF} =$ 15,017 CPU-seconds/simulation), so the difference in $t_\text{PF} - t_\text{ML}$ across the ML models is miniscule. Nonetheless, positive slopes for all lines signify that all seven ML models produce simulations at a faster rate than ParFlow-CLM. The ViT4d technically exhibits the steepest slope, producing simulations the fastest ($t_\text{ML} =0.115$ GPU-seconds/simulation), but, due to the large training set size required for the ViT4d to achieve $<$10\% MAPE, time savings only appear for the ViT4d after 246 simulations. Closer examination of the $x$-intercepts reveals that 6 out of 7 of the ML models achieve a runtime advantage over ParFlow-CLM when the desired simulation output count is as low as 17. 2 out of the 7, the U-FNO4d and PredRNN++, achieve a runtime advantage on and after the fourth simulation. We provide a detailed overview of all time measurements in Table \ref{tab:res_times}, including distinct values of $\mathcal D_\text{train}$ (the minimum size training set required to achieve $<10\%$ MAPE), which strongly impacts the computational efficiency of each model.

When accounting for both training time and forward simulation time, the U-FNO4d is the most efficient ML model. To perform 500 simulations, the U-FNO4d requires a total of 13.1 processor-hours (the sum of Equations \ref{eqn:t_ml} and \ref{eqn:trainingset}), compared to the second-most efficient ML model, the PredRNN++, which requires 13.3 processor-hours. These computational times are orders of magnitude smaller than the time it would take to generate 500 simulations exclusively with ParFlow-CLM (2090 processor-hours). These findings, expressed in Figure \ref{fig:runtime_inset}, affirm the tangible benefits of this hybrid framework, which outpaces a naive approach of solely relying on ParFlow-CLM. We believe this computational speedup arises from two factors. First, ParFlow-CLM uses an iterative Newton-Krylov approach to solve the coupled system of nonlinear partial differential equations at each time step. This iterative scheme requires many Jacobian evaluations and floating point operations per time step, while the ML surrogates only require forward passes, consisting of fixed calculations determined by the model architecture and model weights. Second, the ML surrogate models can take arbitrarily large time steps without compromising convergence due to their data-driven nature, while ParFlow-CLM must use an hourly time step to accurately represent meteorology-driven water fluxes across the land surface. Other process-based models are subject to similar constraints --- such as the Courant–Friedrichs–Lewy condition --- which limit their computational efficiency relative to ML surrogates. Since ML surrogates can learn temporal patterns in training data, they are unbound by time step size and can thereby complete simulations at faster rates.

\begin{table}
\begin{threeparttable}
    \caption{Runtime measurements and GPU memory allocation for all seven ML surrogate models.}
    \begin{tabular}{l c c c c c}
        \toprule
        Model & $t_\text{ML}$ (GPU-sec)\tnote{1} & $T_\text{train}$ (GPU-hr)\tnote{2} &
        $|\mathcal D_\text{train}|$\tnote{3} & $T_\text{training set}$ (CPU-hr)\tnote{4}
        & Memory allocation (GB)\tnote{5} \\
        \midrule
        CNN4d & $2.28$  & $4.47$  & $15$ (13-2) & 62.6 & $21.8$ \\
        ViT4d & $0.0934$  & $1.74$  & $245$ (217-28) & 1020 & $4.2$ \\
        U-FNO4d & $0.856$  & $0.454$  & $3$ (2-1) & 12.5 & $22.7$ \\
        CNN3d & $0.991$  & $0.145$  & $5$ (4-1)& 20.9 & $10.5$ \\
        ViT3d & $0.115$  & $0.122$  & $13$ (11-2)& 54.2 & $2.1$ \\
        U-FNO3d & $0.288$  & $0.187$  & $10$ (8-2)& 41.7 & $14.7$ \\
        PredRNN++ & $2.38$  & $0.473$  & $3$ (2-1)& 12.5 & $31.7$ \\
        \bottomrule
    \end{tabular}
    \begin{tablenotes}
    \begin{footnotesize}
        \item[1] Computational time to perform a single simulation with the trained ML model.
        \item[2] Time to train each ML model, once a training set has been generated.
        \item[3] The minimum number of training examples required to achieve $<$$10\%$ test MAPE. We show the training-validation split in parentheses.
        \item[4] Time to generate $|\mathcal D_\text{train}|$ training examples using the process-based hydrologic model.
        \item[5] Maximum memory allocated to each ML model during training, using a batch size of 1.
    \end{footnotesize}
    \end{tablenotes}
    \label{tab:res_times}
\end{threeparttable}
\end{table}


In addition to runtime, the memory required to train an ML model is an important component of computational efficiency. While the memory cells in the PredRNN++ allow the model to retain information from one time step to the next, these large tensors require significant memory allocation (Table \ref{tab:res_times}). Using a batch size of 1, the PredRNN++ requires 15.1$\times$ more GPU memory to train (31.7 GB) than the ViT3d (2.1 GB), the least memory-intensive architecture. At this allocation size, the PredRNN++ may be too memory-intensive to train on some modern GPUs. Similarly, all three non-sequential models (CNN4d, ViT4d and U-FNO4d) require more memory than their one-step counterparts (CNN3d, ViT3d and U-FNO3d), though this increase in memory allocation does not necessarily translate to an increase in accuracy (Figure \ref{fig:mape_all_stages}). Ultimately, even though some architectures, such as the PredRNN++, can achieve low MAPE with only a few training examples, they may not be flexible enough to adapt to all compute environments due to their high memory requirements.



\subsection{Model interpretability}\label{subsec:interpret}

\begin{figure}
    \centering
    \includegraphics[width=\linewidth]{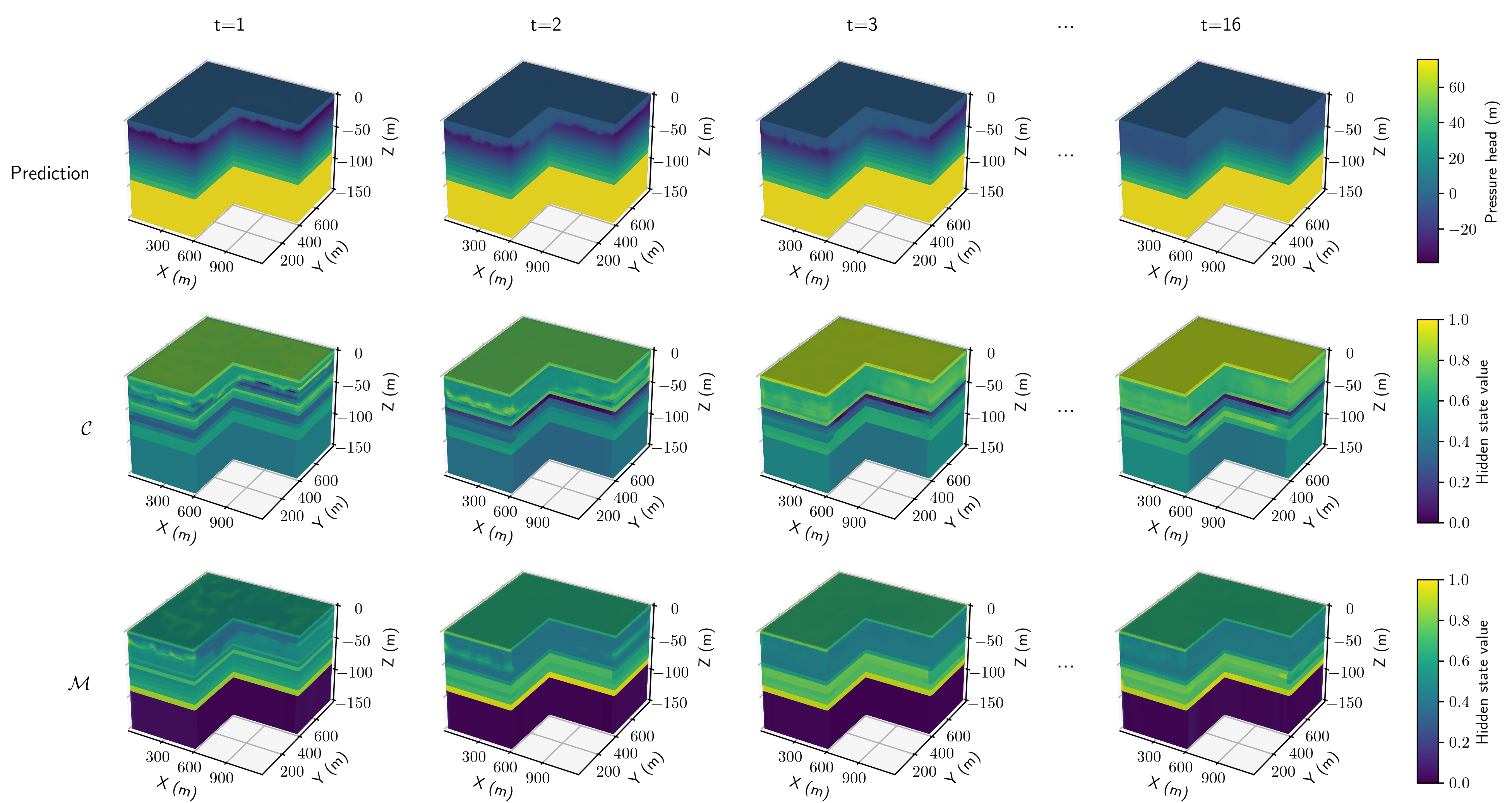}
    \caption{Pressure head (top row) output by the PredRNN++ at three time steps in Stage 1 for a random held-out test example. The $\mathcal C$ memory cells (middle row) and $\mathcal M$ memory cells (bottom row) for the final causal LSTM layer are also shown (see Figure \ref{fig:arch_predrnn} for an overview of the PredRNN++ architecture). Values within $\mathcal C$ and $\mathcal M$ are normalized between 0 and 1 and averaged over the hidden dimension to enable convenient plotting. While $\mathcal{C}$ is intended to store temporal information and $\mathcal M$ is intended to store spatial information, interpretation of these memory cells is difficult because of their size and because local interactions between cells may be difficult to disentangle.}
    \label{fig:interpret}
\end{figure}

Another crucial factor to consider when selecting an ML architecture within a hybrid modeling framework is model interpretability. Error recognition is more straightforward with an interpretable surrogate model and can enhance trust among practitioners, ensuring that a surrogate model is behaving as expected. The PredRNN++ architecture, for example, contains $\mathcal C$ and $\mathcal M$ memory cells (Figure \ref{fig:arch_predrnn}) that are intended to serve distinct roles, with the former facilitating the transfer of temporal information and the latter facilitating the transfer of spatial information \citep{wang_predrnn_2018}. However, in practice, deciphering the values stored in these cells and validating their role in each model has proven challenging (Figure \ref{fig:interpret}). Each memory cell is the same size as the model domain ($120 \times 80 \times 25$), with two memory cells for each of the four causal LSTM layers in the PredRNN++. Given that these memory cells change between each time step and between each evaluation example, manually inspecting and interpreting $\mathcal C$ and $\mathcal M$ quickly becomes cumbersome. In other words, the hidden states of the PredRNN++ model and, consequently, the internal mechanisms of the model, prove challenging to interpret.

Other ML architectures have been shown to be more interpretable than the PredRNN++ when applied in different domains and may prove useful in hydrologic surrogate modeling, such as the CNN3d, which employs convolutional layers, capable of learning hierarchical image information \citep{zhou_object_2014, yang_semantic_2019}, or the ViT3d, whose attention mechanisms can also learn hierarchical information \citep{jain_attention_2019, jawahar_what_2019, wiegreffe_attention_2019}. Future work is needed to explore the interpretablility of each of these architectures when applied to hydrologic simulations.

\subsection{Limitations}\label{subsec:limits}

While each ML model can reliably reproduce transient pressure fields generated by a process-based model, future researchers should carefully tune models to limit spatial and temporal variability in model performance. For example, when loss is calculated evenly across the entirety of the domain, ML models may not accurately simulate changes in pressure head in the most important portions of the domain. In Figure \ref{fig:one_ex}, which visualizes one time step from a PredRNN++ emulation of Stage 1, model errors are predominantly concentrated at and above the wetting front. Even though the loss function is weighted to more severely penalize pressure changes within the vadose zone (Figure \ref{fig:relative_weights}), the high MAPE in this area suggests that the ML model still cannot fully resolve sharp hydraulic gradients near the wetting front. Future applications could employ more complex weighted loss functions to ensure that each surrogate most accurately simulates the processes of interest.

The full results that display ML model performance on all stages in Figure \ref{fig:mape_all_stages} reveal a similar challenge in capturing temporal changes in pressure head. The performance in Stage 1 for all models consistently lags behind that of Stage 2 and Stage 3. Because these later stages simulate shorter time periods (one year in Stage 2 and two years in Stage 3, as opposed to 440 years in Stage 1), pressure head changes over the model stage are less pronounced and their temporal relationships are thereby easier to learn. Thus, while this case study achieves low MAPE, future applications should carefully adapt the implementation strategies in this study (for example, the choice of loss function, data normalization technique, and input features) to their problem of interest, especially when simulating model domains characterized by abrupt variations in pressure head across either time or space.

Finally, while we have examined the impact of modeling choices such as loss function and data normalization across several surrogate architectures, our analysis has been limited to a single case study. Future research should apply these techniques to a broader range of hydrologic scenarios to further validate and generalize these findings.

\section{Conclusion}

Machine learning surrogate models offer an efficient approach for generating large sets of hydrologic simulations. All seven of the ML surrogate models produce transient pressure fields that are comparable (below 10\% mean absolute percentage error, MAPE) to those produced by the process-based hydrologic model. These transient pressure fields can then be used in downstream tasks, such as quantifying changes in groundwater storage. The improved predictive recurrent neural network (PredRNN++) exhibits the highest accuracy when simulating all three model stages (0.36\% MAPE), though other architectures also perform well (0.52--2.4\% MAPE). To achieve $<$10\% MAPE for this case study, some architectures such as the U-FNO4d and PredRNN++ require $<$10 training examples.

This hybrid modeling framework yields substantial runtime benefits, though the threshold at which it becomes more efficient than a process-based model varies by architecture. Once trained, each architecture requires 0.115--2.38 processor-seconds to generate a single Stage 1 simulation, as opposed to the process-based model, which requires 3,860--69,700 processor-seconds/simulation. When accounting for the time it takes to train each model, 6 out of 7 ML surrogate models (all except for the four-dimensional vision transformer) exhibit runtime advantages when generating as few as 17 hydrologic simulations. Because they require so few training examples, the U-FNO4d and PredRNN++ achieve runtime advantages on and after the fourth simulation. These findings indicate that, when generating more than 17 hydrologic simulations for this case study, a surrogate model is more computationally efficient than a process-based model, even when including the time required to generate a training set and train the surrogate. In contrast, when performing fewer than four simulations for this case study, a process-based model alone is more efficient.

Additionally, we identify the impacts of key modeling choices on surrogate accuracy, showing that min-max data normalization can improve surrogate model accuracy by up to a factor of 10 when compared to other treatments such as Z-score and no normalization. Selecting a normalized loss function also improves training stability. Downsampling input features using an autoencoder, meanwhile, reduces memory requirements by training with tensors 4\% of their original size.

Overall, this study offers encouraging results for applying machine learning surrogate models to real-world hydrologic problems, supporting future advancements in hydrologic modeling and simulation techniques.

\subsubsection*{Acknowledgements}
We thank the Stanford Research Computing Center for providing computational resources and support that contributed to these research results. This material is based upon work supported by the National Science Foundation Graduate Research Fellowship under Grant No. DGE-1656518. Additional research funding was provided by the Realizing Environmental Innovation Program (REIP) at the Stanford Woods Institute for the Environment.

All process-based hydrologic simulations were performed with ParFlow-CLM v3.10.0 \citep[\url{https://doi.org/10.5281/zenodo.6413322}; ][]{smith_parflow_2022}. Output files from process-based hydrologic simulations are available to the public via the Stanford Digital Repository \citep[\url{https://doi.org/10.25740/hj302gv2126}; ][]{perzan_datamodel_2023}. All code used to implement, train and evaluate each machine learning surrogate architecture is available to the public in HydroShare \citep[\url{https://doi.org/10.4211/hs.f0a31fbc3de148a98deb36795b4fac53}; ][]{dai_code_2025}.

\clearpage
\begin{appendix}

\setcounter{section}{0}
\setcounter{table}{0}
\setcounter{figure}{0}
\renewcommand{\thesection}{S\arabic{section}}
\renewcommand{\thetable}{S\arabic{table}}
\renewcommand{\thefigure}{S\arabic{figure}}

\section*{Overview}
This appendix contains additional details on the methods used to develop each ML surrogate model, including layer-by-layer information for each surrogate model (\S \ref{appendix:arch}), diagrams of recent model architecture innovations (\S \ref{appendix:arch_diags}), and details on the autoencoder used to compress 3D pressure fields (\S \ref{appendix:autoenc}). We also present additional results referenced in the main text, including complete MAPE results across all models and stages (\S \ref{appendix:mape_all_stages}) and a discussion of any error introduced by the autoencoder (\S \ref{appendix:autoenc_error}).

\section{Materials and methods}

\subsection{Model layer tables}\label{appendix:arch}

\begin{spacing}{1.1}
    \begin{longtable}{r l l l }
         & Layer& Output shape & Parameter count\\
        \hline
        1. & Linear & $(-1,120,80,25,64)$ & 576\\
        2. & Conv3d/BatchNorm3d/ReLU/Dropout & $(-1,64,120,80,25)$ & 512,192\\
        3. & Conv3d/BatchNorm3d/ReLU/Dropout & $(-1,128,120,80,25)$ & 1,024,384\\
        4. & Conv3d/BatchNorm3d/ReLU/Dropout & $(-1,64,120,80,25)$ & 1,024,384\\
        5. & Linear & $(-1,120,80,25,1)$ & 65\\
    \caption{CNN3d layer details given $(-1, 120, 80, 25, 8)$ input.}
    \end{longtable}

    \begin{longtable}{r l l l }
         & Layer& Output shape & Parameter count\\
        \hline
        1. & Linear & $(-1,1,120,80,25,64)$ & 576\\
        2. & Conv4d/BatchNorm4d/ReLU/Dropout & $(-1,64,1,120,80,25)$ & 2,560,128\\
        3. & Conv4d/BatchNorm4d/ReLU/Dropout & $(-1,128,16,120,80,25)$ & 5,120,256\\
        4. & Conv4d/BatchNorm4d/ReLU/Dropout & $(-1,64,16,120,80,25)$ & 5,120,256\\
        5. & Linear & $(-1,16,120,80,25,1)$ & 65\\
    \caption{CNN4d layer details given $(-1, 16, 120, 80, 25, 8)$ input.}
    \end{longtable}

    \begin{longtable}{r l l l}
        & Layer& Output shape & Parameter count\\
        \hline
        1. & Padding & $(-1,120,80,32,8)$ & 0\\
        2. & Linear & $(-1,120,80,32,32)$ & 256\\
        3. & Linear & $(-1,120,80,32,32)$ & 1056\\
        4. & Fourier3d/Conv1d/Add/ReLU & $(-1,32,120,80,32)$ & 4,097,056\\
        5. & Fourier3d/Conv1d/Add/ReLU & $(-1,32,120,80,32)$ & 4,097,056\\
        6. & Fourier3d/Conv1d/Add/ReLU & $(-1,32,120,80,32)$ & 4,097,056\\
        7. & Fourier3d/Conv1d/U-Net3d/Add/ReLU & $(-1,32,120,80,32)$ & 4,618,720\\
        8. & Fourier3d/Conv1d/U-Net3d/Add/ReLU & $(-1,32,120,80,32)$ & 4,618,720\\
        9. & Fourier3d/Conv1d/U-Net3d/Add/ReLU & $(-1,32,120,80,32)$ & 4,618,720\\
        10.&Linear&$(-1,120,80,32,128)$ & 4224\\
        11.&Linear&$(-1,120,80,32,1)$ & 129\\
        12.&Depadding&$(-1,120,80,25,1)$ & 0\\
        \caption{UFNO3d layer details given $(-1, 120, 80, 25, 8)$ inputs.}
    \end{longtable}

    \begin{longtable}{r l l l}
        & Layer& Output shape & Parameter count\\
        \hline
        1. & Padding & $(-1,16,120,80,32,8)$ & 0\\
        2. & Linear & $(-1,16,120,80,32,32)$ & 256\\
        3. & Fourier4d/Conv1d/Add/ReLU & $(-1,32,16,120,80,32)$ & 40,961,056\\
        4. & Fourier4d/Conv1d/Add/ReLU & $(-1,32,16,120,80,32)$ & 40,961,056\\
        5. & Fourier4d/Conv1d/Add/ReLU & $(-1,32,16,120,80,32)$ & 40,961,056\\
        6. & Fourier4d/Conv1d/U-Net4d/Add/ReLU & $(-1,32,16,120,80,32)$ & 42,852,832\\
        7. & Fourier4d/Conv1d/U-Net4d/Add/ReLU & $(-1,32,16,120,80,32)$ & 42,852,832\\
        8. & Fourier4d/Conv1d/U-Net4d/Add/ReLU & $(-1,32,16,120,80,32)$ & 42,852,832\\
        9.&Linear&$(-1,16,120,80,32,128)$ & 4224\\
        10.&Linear&$(-1,16,120,80,32,1)$ & 129\\
        11.&Depadding&$(-1,16,120,80,25,1)$ & 0\\
        \caption{UFNO4d layer details given $(-1, 16, 120, 80, 25, 8)$ inputs.}
    \end{longtable}

    \begin{longtable}{r l l l}
        & Layer& Output shape & Parameter count\\
        \hline
        1. & Linear & $(-1,120,80,25,64)$ & 512\\
        2. & Conv3d & $(-1,64, 5, 5, 5)$ & 7,864,320\\
        3. & PositionalEncoding & $(125, -1, 64)$ & 0\\
        4. & TransformerEncoder & $(125, -1, 64)$ & 2,249,216\\
        5. & Linear & $(-1, 125, 1920)$ & 124,800\\
        6. & Reshape & $(-1, 120, 80, 25, 1)$ & 0\\
        \caption{ViT3d layer details given $(-1, 120, 80, 25, 8)$ inputs.}
    \end{longtable}

    \begin{longtable}{r l l l}
        & Layer& Output shape & Parameter count\\
        \hline
        1. & Linear & $(-1,16, 120,80,25,64)$ & 512\\
        2. & Conv4d & $(-1,64, 8, 5, 5, 5)$ & 15,728,704\\
        3. & PositionalEncoding & $(1000, -1, 64)$ & 0\\
        4. & TransformerEncoder & $(1000, -1, 64)$ & 2,249,216\\
        5. & Linear & $(-1, 1000, 3840)$ & 249,600\\
        6. & Reshape & $(-1, 16, 120, 80, 25, 1)$ & 0\\
        \caption{ViT4d layer details given $(-1, 16, 120, 80, 25, 8)$ inputs.}
    \end{longtable}

    \begin{longtable}{r l l l}
        & Layer& Output shape & Parameter count\\
        \hline
        1. & Linear & $(-1, 120,80,25,32)$ & 256\\
        2. & CausalLSTM3d & $(-1,32, 120, 80, 25)$ & 2,690,752\\
        3. & GHU & $(-1,32, 120, 80, 25)$ & 512,128\\
        4. & CausalLSTM3d & $(-1,32, 120, 80, 25)$ & 2,690,752\\
        5. & CausalLSTM3d & $(-1,32, 120, 80, 25)$ & 2,690,752\\
        6. & CausalLSTM3d & $(-1,32, 120, 80, 25)$ & 2,690,752\\
        7. & Linear & $(-1,120, 80, 25, 1)$ & 33\\
        \caption{PredRNN++ layer details given $(-1, 120, 80, 25, 8)$ inputs.}
    \end{longtable}
\end{spacing}

\subsection{Model architecture diagrams}\label{appendix:arch_diags}

\subsubsection{U-FNO}

\begin{figure}[H]
    \centering
    \includegraphics[width=\textwidth]{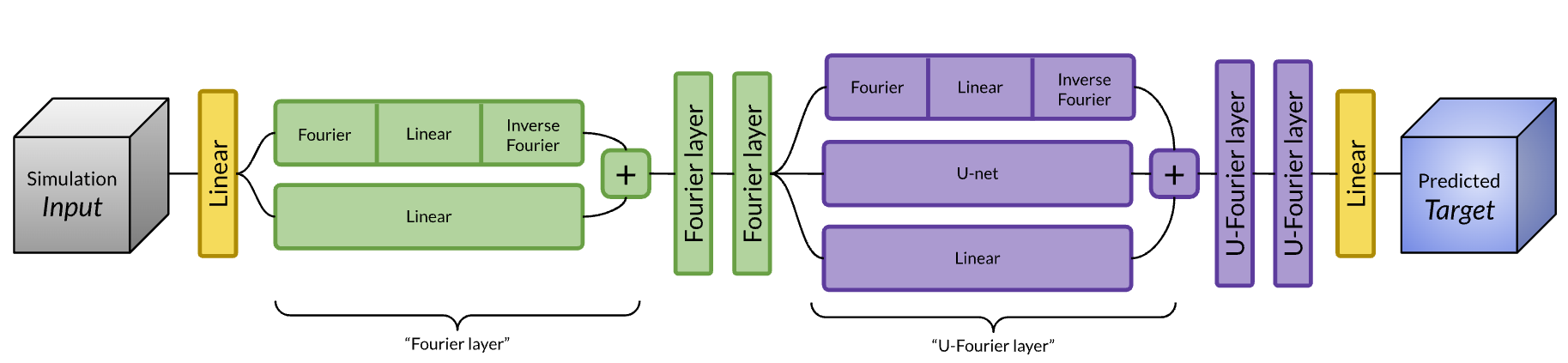}
    \caption{U-FNO \citep{wen_u-fnoenhanced_2022} diagram.}
    \label{fig:arch_ufno}
\end{figure}

\subsubsection{PredRNN++}

\begin{figure}[H]
    \centering
    \includegraphics[width=\textwidth]{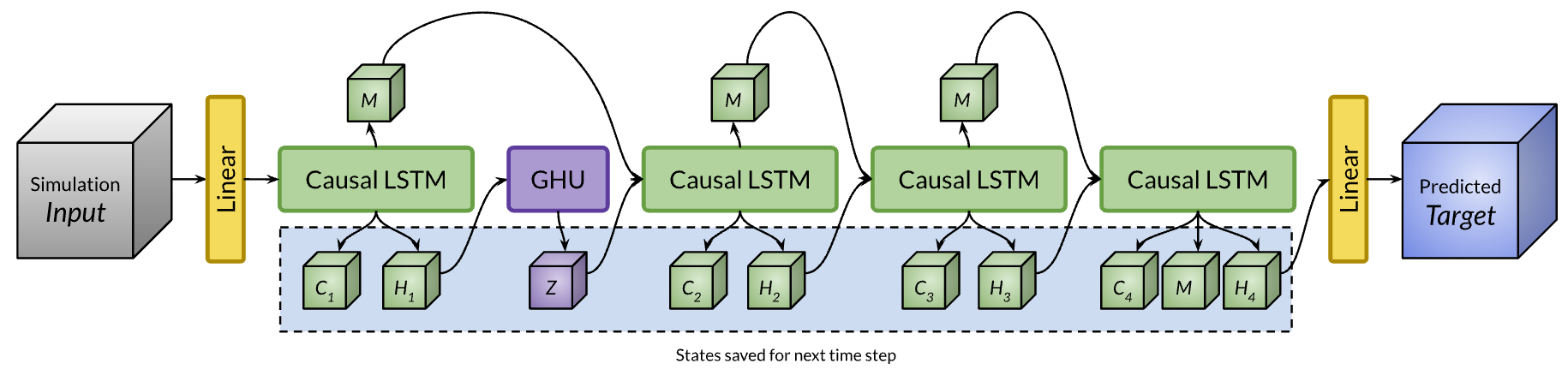}
    \caption{PredRNN++ \citep{wang_predrnn_2018} architecture diagram. The memory cells in the blue highlighted region are saved for the next time step. As the forward pass for the next time step is being computed, the model will utilize the saved memory cells from the previous forward pass. Note that the temporal $\mathcal C$ memory is not passed to deeper layers within the model; it is retained within the same causal LSTM layer across each time step.}
    \label{fig:arch_predrnn}
\end{figure}

\subsection{Autoencoder details} \label{appendix:autoenc}
The autoencoder is a convolutional neural network containing four layers and around 676,000 parameters, which makes the autoencoder much smaller than each surrogate model. Two of the four layers are encoding convolutional layers, which squeezes the input into compact spatial dimensions, and the remaining two are decoding transposed convolutional layers, which recovers the original spatial resolution. All convolutional layers are 2D, using the $z$ dimension as the channel dimension. They employ a (7, 7)-sized convolutional kernel, a stride of (2, 2), and a hidden size of 128.  The convolutional layers are separated by ReLU activation functions and batch normalization layers.

\subsection{Implicit depth-wise loss and MAPE weighting}\label{appendix:rel_weight}

\begin{figure}[H]
    \centering
    \includegraphics[width=3in]{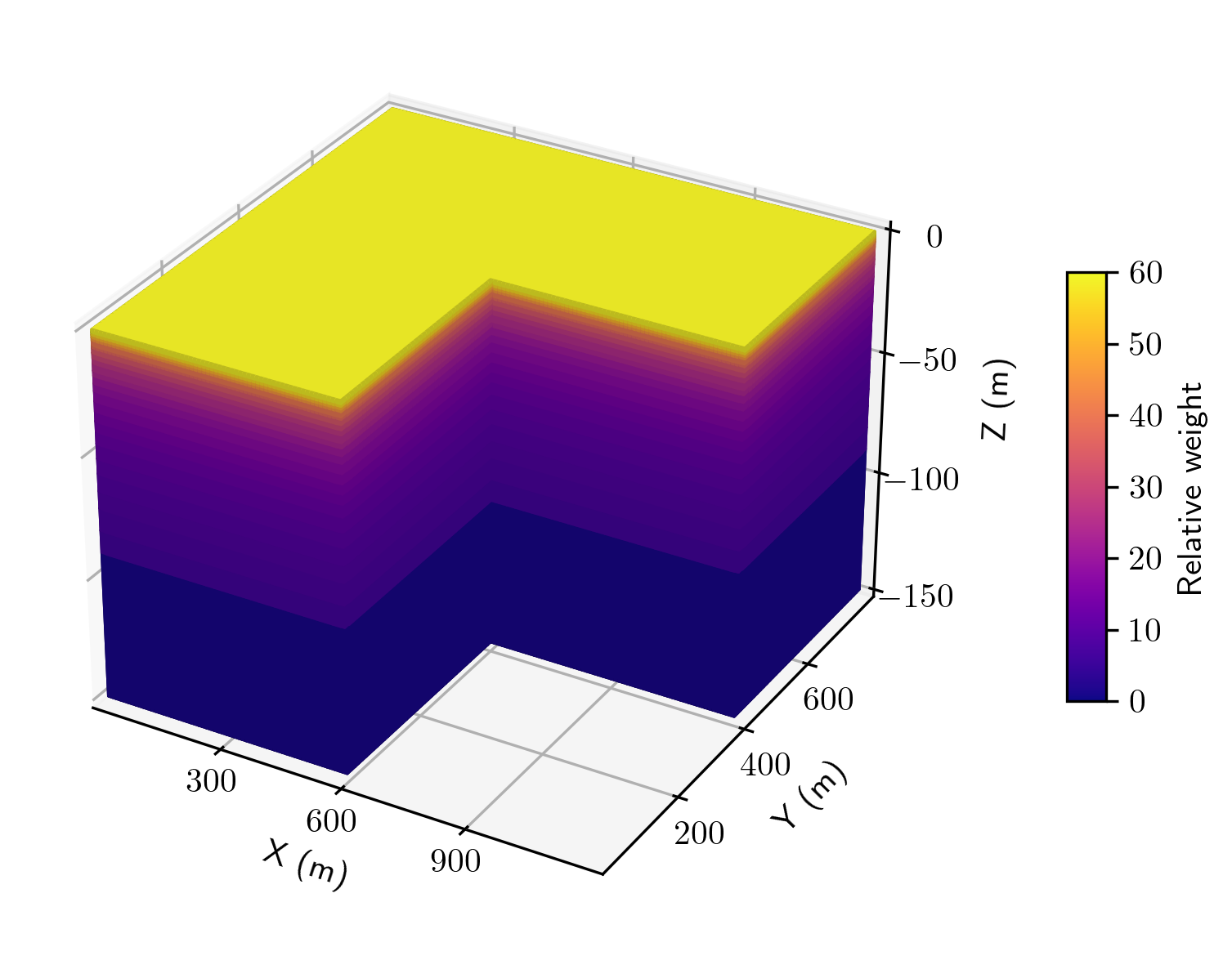}
    \caption{Layers near the surface of the rectilinear grid are implicitly weighted higher than deeper layers in loss calculations and MAPE calculations. This is due to the fact that layers have variable real-world heights: the deepest layer is 60$\si{m}$, and each successive layer is a fraction of the depth of this layer.}
    \label{fig:relative_weights}
\end{figure}

\section{Results}

\subsection{Mean absolute percentage error across all model stages}
\label{appendix:mape_all_stages}
\begin{table}[H]
    \centering
    \centerline{
    \begin{tabular}{c l c c c c}
        \toprule
        & & \multicolumn{4}{c}{MAPE}\\
        \cmidrule{3-6}
        & Model & Stage 1 & Stage 2 & Stage 3 & E2E\\
        \midrule
a. & CNN4d & $5.4$  & $0.089$  & $0.75$  & $2.4$\\
b. & ViT4d & $9.9$  & $0.14$  & $0.28$& $0.70$\\
c. & U-FNO4d & $4.3$  & $0.14$  & $0.28$& $0.52$\\
d. & CNN3d & $4.8$  & $\mathbf{0.050}$  & $0.29$& $1.9$  \\
e. & ViT3d & $6.3$  & $0.12$  & $\mathbf{0.20}$& $0.57$\\
f. & U-FNO3d & $5.1$& $0.22$  & $0.22$& $0.62$\\
g. & PredRNN++ & $\mathbf{2.9}$  & $0.054$  & $0.46$& $\mathbf{0.36}$  \\

        \bottomrule
    \end{tabular}
    }
    \caption{Test MAPEs for all ML models. Bold indicates best performance for each stage.}
    \label{tab:res_compile}
\end{table}

\subsection{Error introduced by the autoencoder}
\label{appendix:autoenc_error}
\begin{figure}
    \centering
    \includegraphics[width=5in]{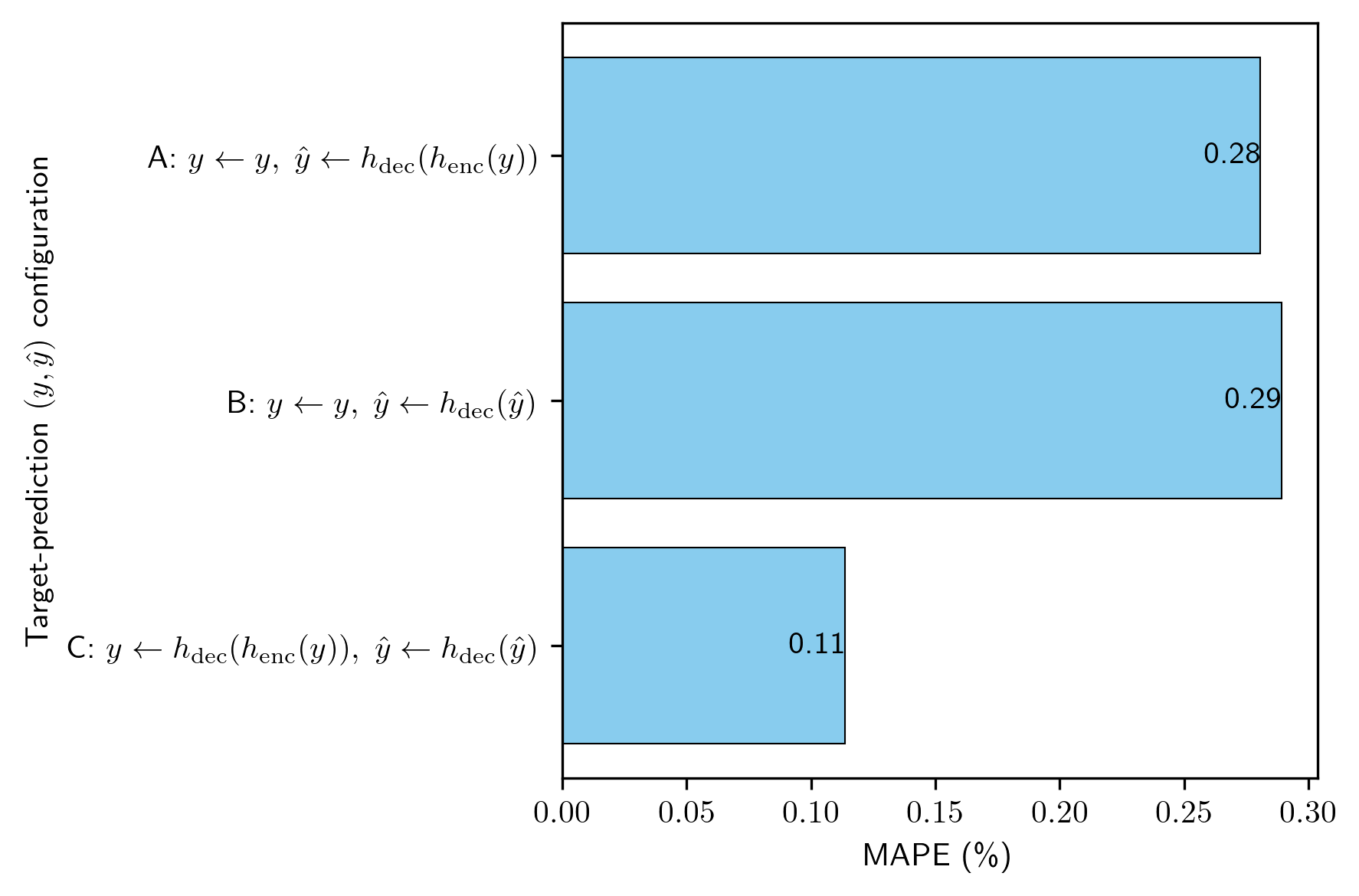}
    \caption{Comparison of autoencoder percentage errors under various target-prediction ($y, \hat y$) configurations. Configuration A represents the autoencoder error. Configuration B represents Stage 3 model error. Configuration C modifies the analysis of Stage 3 model error by substituting the targets with targets that are encoded then decoded with the autoencoder.}
    \label{fig:autoenc_error}
\end{figure}

The autoencoder, though a vital component in training Stage 3 models, may introduce inaccuracies. We attempt to disentangle the errors from the surrogate model and the errors from the autoencoder in Figure \ref{fig:autoenc_error}, focusing on the CNN3d for its efficiency in training. Computing MAPE between the predictions $\hat y$ taken directly from the autoencoder, i.e., $\hat y \leftarrow h_\text{dec}(h_\text{enc}(y))$, and the simulation targets reveals that the autoencoder can reproduce its own input with 0.28\% error on the held-out test set (Figure \ref{fig:autoenc_error}, Config A). We expect that this autoencoder error contributes to Stage 3 error, since we use the autoencoder in training Stage 3 models; indeed, when computing the MAPE between decoded Stage 3 model predictions and simulation targets, this error, at 0.29\%, is greater than the autoencoder error alone. It is surprising, then, when we use the autoencoder to encode and decode the targets and use these encoded and decoded targets as our new evaluation references, our percent error decreases to 0.11\%. This shift indicates that the autoencoder introduces error by compressing the space of the target encodings. In other words, our autoencoder encodes targets in a way that makes it more difficult to distinguish between unique targets once they have been encoded. Consequently, during the decoding step, essential information distinguishing unique targets is not fully recovered. Exploring improved autoencoder architectures may help to mitigate these challenges and enhance the overall performance our hybrid framework in future applications.

\end{appendix}

\clearpage
\bibliographystyle{unsrtnat}
\bibliography{biblio}  






\end{document}